\documentclass{emulateapj}
\usepackage[breaklinks,colorlinks,citecolor=blue]{hyperref}

\usepackage{natbib}
\usepackage{threeparttable}
\usepackage{longtable}
\usepackage{epstopdf}
\usepackage{color}
\usepackage[all]{hypcap}

\slugcomment{Accepted to Publications of the Astronomical Society of the Pacific}

\shorttitle{NSB at OAN-SPM}
\shortauthors{Plauchu-Frayn et al.}

\begin{document}

\title{Night sky brightness at San Pedro Martir Observatory}

\author{I. Plauchu-Frayn\altaffilmark{1},  M.~G. Richer\altaffilmark{1}, E. Colorado\altaffilmark{1}, 
J. Herrera\altaffilmark{1}, A. C\'ordova\altaffilmark{1}, U. Cese\~na\altaffilmark{1}, F. \'Avila. \altaffilmark{1}}

\altaffiltext{1}{Instituto de Astronom\'\i a, Universidad Nacional Aut\'onoma de M\'exico, 
Apartado Postal 106, 22800 Ensenada, B.C., M\'exico. }

\begin{abstract}

We present optical UBVRI zenith night sky brightness measurements collected on eighteen nights during 2013--2016 and SQM 
measurements obtained daily over twenty months during 2014--2016 at the Observatorio Astron\'omico 
Nacional on the Sierra San Pedro M\'artir (OAN-SPM) in M\'exico. The UBVRI data is based upon CCD images obtained with 
the 0.84\,m and 2.12\,m telescopes, while the SQM data is obtained with a high-sensitivity, low-cost photometer. 
The typical moonless night sky brightness at zenith averaged over the whole period is U = 22.68, B = 23.10, V = 21.84, 
R = 21.04, I = 19.36, and SQM = 21.88 $\mathrm{mag\,arcsec^{-2}}$, once corrected for zodiacal light. We find no 
seasonal variation of the night sky brightness measured with the SQM. The typical night sky brightness values found at 
OAN-SPM are similar to those reported for other astronomical dark sites at a similar phase of the solar cycle. We find a 
trend of decreasing night sky brightness with decreasing solar activity during period of the observations. This trend 
implies that the sky has become darker by $\Delta U =$0.7, $\Delta B =$0.5, $\Delta V =$0.3, $\Delta R =$0.5 mag arcsec$^{-2}$ 
since early 2014 due to the present solar cycle.
\end{abstract}


\section{Introduction}

The Observatorio Astron\'omico Nacional San Pedro M\'artir (hereafter OAN-SPM) is located on the top of Sierra San Pedro 
M\'artir in Baja California, M\'exico (2800\,m, +31$^{\circ}$\,02''\, 40'\, N, 115$^{\circ}$\,28''\, 00'\, W). 
The site excels in sky clarity with, in recent decades, approximately 70\% and 80\% photometric and spectroscopic time, 
respectively \citep{2007RMxAC..31...47T}. The median seeing measured at zenith at 5000\AA\ varies from 0.''50 to 0''.79
(\citealp{1998RMxAA..34...47E}; \citealp{2003RMxAC..19...37M}; \citealp{2009PASP..121.1151S}; \citealp{2012MNRAS.426..635S}). 
Atmospheric extinction is typically 0.13\,mag airmass$^{-1}$ in V band (\citealp{2001RMxAA..37..187S}). Due to these 
excellent atmospheric conditions and favorable location away from large urban areas, the OAN-SPM is an excellent 
site for optical and infrared facilities. \\

Among the most important parameters that define the quality of an observing site it is the night sky brightness (NSB). This 
parameter has been extensively studied by several authors (\citealp{1975PASP...87..869K}; \citealp{1988PASP..100..496W}; 
\citealp{1989PASP..101..707P}; \citealp{1987PASP...99..887K}; \citealp{1995A&AS..112...99L}; 
\citealp{1996A&AS..119..153M}; \citealp{2003A&A...400.1183P} and references therein), starting with the pioneering 
work by \citet{1928RSPSA.119...11R}. In the following, we will concentrate on  optical wavelengths only.\\

The NSB is the integrated light from two main kinds of sources: natural and artificial. Among 
the sources of natural origin are airglow (recombination of molecules heated by Sun UV radiation during daytime), 
aurorae, zodiacal light (sunlight scattered from interplanetary dust), diffuse galactic light (from faint unresolved 
stars in our Galaxy), and the extragalactic background (due to distant, faint unresolved galaxies).  The airglow and 
aurorae, which originate in the Earth's atmosphere, depend upon the site and time of the observation, while the 
other three do not. The source of artificial light is mainly street lighting, with an increasing contribution 
from electronic billboards and other luminous advertising media.  This  contribution, also known as light pollution, 
is amenable to monitoring through long-term campaigns of the variation in the night sky brightness (\citealp{1979PASP...91..530S}; 
\citealp{1988PASP..100..496W}; \citealp{1975PASP...87..869K}; \citealp{1987PASP...99..887K}; \citealp{1989PASP..101..707P}; 
\citealp{1995A&AS..112...99L}).\\

\begin{figure*}
 \begin{center}
 \includegraphics[width=0.80\textwidth]{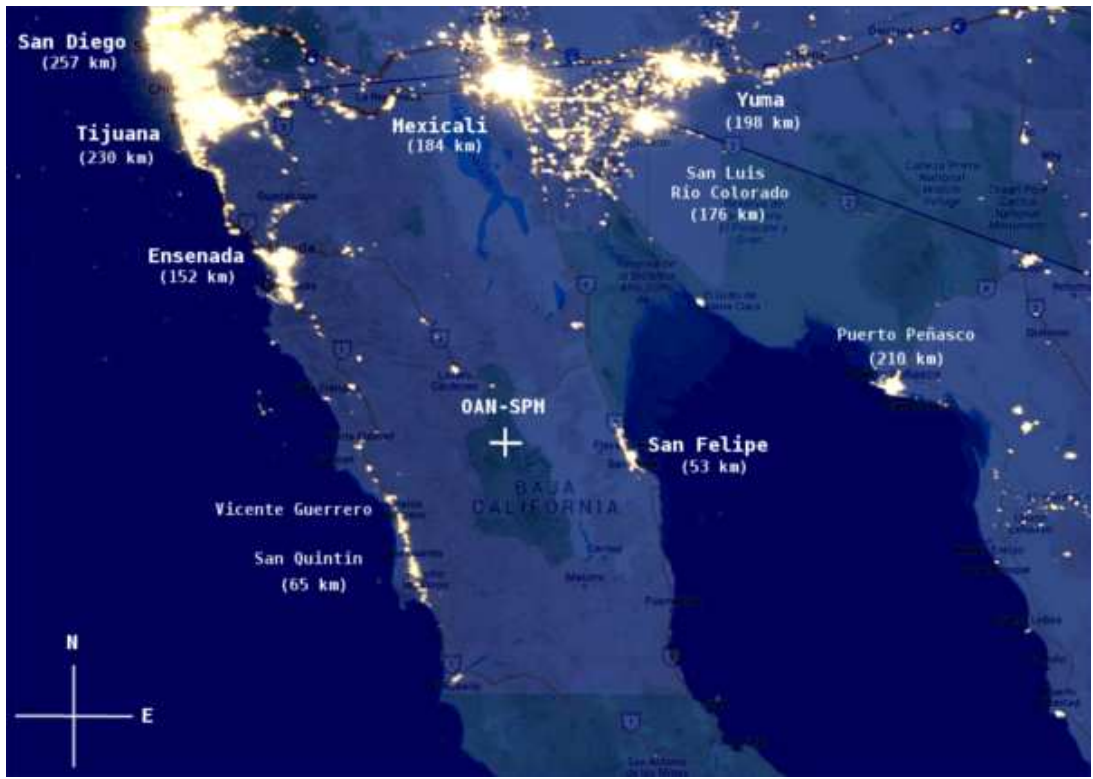}
   \caption{Night map showing nearby cities and their approximate distances to OAN-SPM. The cross indicates the 
   location of the OAN-SPM. Bright areas indicate street lights as seen from space. Credit: Google maps and NightEarth.com
   with data provided by NASA.}  \label{fig:cities}
\end{center}
\end{figure*}

In Fig.~\ref{fig:cities} we show the cities and towns near the OAN-SPM. The cities of Ensenada and Tijuana lie between 
150 and 230\,km NW of OAN-SPM and have populations of 480,000 and 1.6 million people, 
respectively. The city of San Diego lies 260\,km distant, also to the NW, with a population of 1.3 million people. 
An estimate of the contribution to the NSB due to light from nearby cities can be obtained using the 
model of \citet{1989PASP..101..306G}, which provides an approximate light-pollution contribution expected from different 
sources. The combined contribution of Ensenada, Tijuana and San Diego to the 
sky brightness at the OAN-SPM is estimated to be less than 0.08\,mag at a zenith distance of 45$^{\circ}$.  
To the Northeast, at an average distance of 185\,km, the cities of Mexicali, Yuma, and San Luis R\'\i o Colorado, 
with a combined population of 1.3 million contribute with 0.04\,mag.  Other cities like San Felipe and San Quint\'\i n, 
which are nearer to the observatory ($\sim$60\,km), but less populated ($\sim$17\,000 and 10\,000 people) contribute  
$<$0.01\,mag each. In Baja California, state lighting ordinances that took effect starting in 2006 in the municipality of 
Ensenada and statewide in 2010 include light pollution among the environmental disturbances to be controlled. 
Among its goals, this legislation seeks to reduce or at least minimize the light pollution, even with the constant 
growth of its cities.\\

In the present paper we report  UBVRI sky brightness measurements  obtained on eighteen moonless nights during 
2013 to 2016 and SQM sky brightness measurements collected daily during 2014 to 2016.  As far as we are aware, these data constitute the 
largest homogeneous data set available for the OAN-SPM. The SQM data set is continuously accumulating and it will 
provide an unprecedented opportunity to investigate the long-term evolution of the night sky at the OAN-SPM.  In 
Sect.~\ref{sec:observ}, we present our observation procedures and reduction techniques. The results are presented 
in Sect.~\ref{sec:results}, while, in Sect.~\ref{sec:discuss}, we consider the variation of the NSB as a function of the solar 
activity and compare our measurements with other dark sites. Finally, in Sect.~\ref{sec:concl}, we 
present our conclusions.

\section{Observations and data reduction}
\label{sec:observ}

\subsection{CCD night sky brightness measurements}
\label{sec:ccdobs}

The broadband data set presented in this study was obtained with the MEXMAN and Italian filter wheels, which are 
mounted at the Cassegrain focus of the 0.84\,m and 2.1\,m Ritchey-Chretien telescopes, respectively.  The detectors are E2V 
back-illuminated CCDs with 13.5\,$\mu$m pixels in a $2048\times 2048$ format, which give projected plate scales 
of 0''.22/pix and 0''.18/pix at the 0.84\,m and 2.1\,m telescopes, respectively.\\

Observations of the NSB were obtained at the zenith to minimize airglow emission and extinction 
on 18 photometric nights between February 2013 to May 2016 when both the Sun and Moon were at least 18$^{\circ}$ below 
the horizon (astronomical night).  Each night, a photometric standard field (\citealp{1992AJ....104..340L}) was 
observed at about the same time as part of the calibration process. \\

Imaging frames are bias and flat-field corrected using standard reduction procedures in IRAF\footnote{IRAF is 
distributed by the National Optical Astronomy Observatory, which is operated by AURA, INC. under cooperative 
agreement with the National Science Foundation}. Average integration times range between 600s in U band and 
300s in I band, in order to obtain sufficient sky counts. Aperture photometry was performed with the APT software 
(\citealp{2012PASP..124..737L}) by using  an aperture of 13$^{\circ}$ in areas free of nebulosity, stars, and 
cosmic rays.  The instrumental magnitudes of the standard stars were corrected for atmospheric extinction, 
using the standard values for the OAN-SPM  (\citealp{2001RMxAA..37..187S}).  No color correction was applied 
to these magnitudes.  Following the prescriptions of \citet{1989PASP..101..707P}, the NSB is calibrated 
without correcting for atmospheric extinction because we are interested in the observed brightness. The typical 
average uncertainty in the photometry, neglecting the uncertainty in the exposure time and aperture size, in 
our NSB measurements are 0.13, 0.05, 0.04, 0.03 and 0.02\,mag for U, B, V, R, and I bands, 
respectively, based upon image statistics. Our complete NSB data set from CCD imaging is tabulated in Table \ref{tab:allnights}. \\

\begin{figure}[h!]
 \centering
 \includegraphics[width=0.45\textwidth]{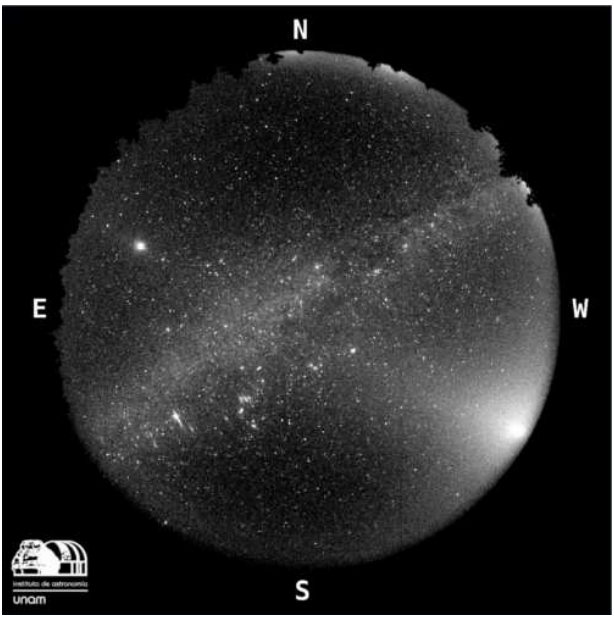}
  \caption{The Milky Way (center) and zodiacal light (SW) are seen clearly in this image taken by the All--sky camera at 
  OAN-SPM on 18 February 2015.}
  \label{fig:mw_zl}
\end{figure}

\begin{figure}[h!]
 \centering
 \includegraphics[width=0.50\textwidth]{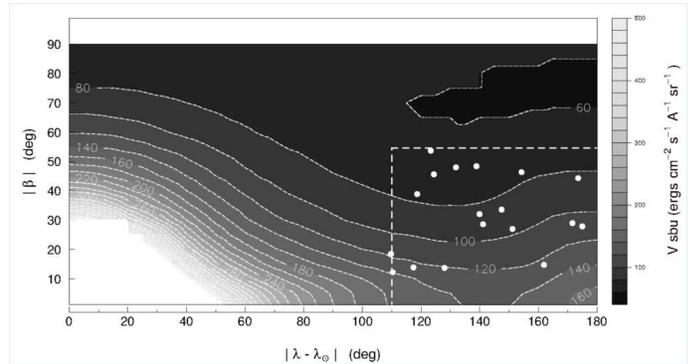}
   \caption{We superpose the distribution of the telescope pointings in helio-ecliptic coordinates (white circles) 
   upon a contour    plot of the zodiacal light in the V-band.  The surface brightness of the zodiacal light 
   is expressed in surface brightness    units, sbu (erg\, s$^{-1}$\, cm$^{-2}$\,\AA$^{-1}$\,sr$^{-1}$). The 
   region inside the box    $|\,\lambda-\lambda_{\odot}\,|>$110$^{\circ}$\, and $|\,\beta\,|< $ 54.5$^{\circ}$\,(dashed line) 
   contains the zenith pointings    of the SQM in helio-ecliptic coordinates. Original data for zodiacal light 
   contours are from \citet{1980A&A....84..277L}.}  
   \label{fig:Zodiacallight_VbandS10}
\end{figure}

The NSB has an important contribution from zodiacal light that has to be taken into account. In 
Fig.~\ref{fig:mw_zl}, we show an image of the all-sky camera installed at the OAN-SPM for the night of 18 
February 2015 where we may appreciate the presence of the Milky Way at the zenith and the zodiacal light 
to the SW on the horizon. In Fig.~\ref{fig:Zodiacallight_VbandS10}, we have superimposed the telescope pointings on a contour plot of 
the zodiacal light V  brightness in helio-ecliptic coordinates (\citealp{1980A&A....84..277L}), i.e., ecliptic latitude versus 
the difference in ecliptic longitude of the observation and that of the Sun. We have used the equatorial celestial 
coordinates at the zenith when the sky brightness was measured and 
converted the right ascension and declination to helio-ecliptic coordinates.  The surface brightness contours are expressed  
in surface brightness units, sbu  (ergs\,s$^{-1}$\,cm$^{-2}$\,\AA$^{-1}$\,sr$^{-1}$), e.g., a typical 
V sky brightness of 21.6 mag arcsec$^{-2}$ is equivalent to 366 sbu.  The spectrum of the 
zodiacal light is very similar to that of the Sun over the UV-IR range, and peaks at 4500\,\AA.  In order to reduce 
the scatter in our NSB measurements, we remove the zodiacal light contribution in the UBVRI passbands for each telescope 
pointing (see Table \ref{tab:allnights} for the individual corrections).  The average  contribution to 
the total NSB is 45\%, 60\%, 25\%, 10\%, and 4\%  in the U, B, V, R, and I bands, respectively (see Table~\ref{tab:nsb}). \\

\subsection{SQM night sky brightness measurements}
\label{sec:sqmobs}

Since 2 November 2014, we have monitored the sky with an Unihedron Sky Quality Meter\footnote{\url{http://www.unihedron.com}} 
(hereafter SQM) in a continuous manner. This is a low cost NSB 
photometer with high enough sensitivity to quantify the quality of the night sky at any place. The SQM is encased in 
a weatherproof housing pointing at the zenith.  Periodically, the housing is cleaned manually. Our study includes 
data spanning 607 nights from November 2014 to June 2016, of which 534 (88\%) have available data.   The SQM has a 
spectral response similar, though not identical, to the V band. Its sensitivity peaks at 5400\,\AA\, with a broad 
transmittance window ($\sim$2000\,\AA; see \citet{Cinzano05} for a detailed comparison with stardard photometric 
systems). Here, we follow the convention of other authors and report all measurements in terms of the SQM spectral 
band unit, mag$_{SQM}$ arcsec$^{-2}$. \\

The SQM measures the NSB every minute in a cone of about 20$^{\circ}$\,(FWHM) and reports the result in astronomical 
units of magnitudes per square arcsecond with a precision of 0.1\,mag arcsec$^{-2}$. Due to the large field of view, 
this device detects both the zenithal and near-zenithal NSB, underestimating the true NSB of the zenith (lower 
values, i.e. brighter magnitudes). This over-estimate is about $0-0.3$\,mag arcsec$^{-2}$, depending upon 
the light pollution at the site. 
Also, the SQM includes the integrated light from all stars within the field of view, which contribute approximately 
6\% of the NSB for stars of magnitude $\le$5\,mag and must be accounted for (\citealp{Cinzano05}). We do not apply 
corrections for either light pollution or the integrated light of stars.\\

\begin{figure}[h!]
 \centering
 \includegraphics[width=0.50\textwidth]{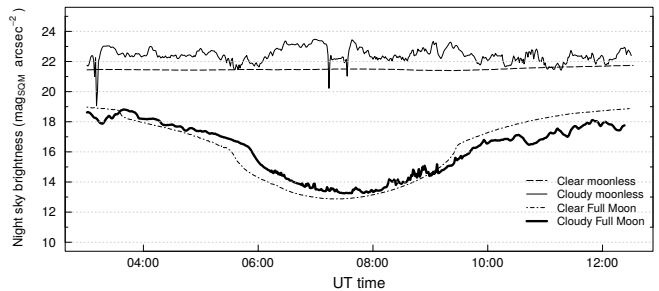}
  \caption{Examples of the variation in the SQM measurements under different conditions of cloud cover. }
  \label{fig:SQM_examples}
\end{figure}

In Fig.~\ref{fig:SQM_examples}, we present the SQM data plotted for four different nights: a cloudy moonless 
night, a clear moonless night, a clear moonlit night, and a cloudy moonlit night, in order to show the expected 
variations of the SQM data. The sky brightness on clear nights ranges from the brightest value of 
13\,mag$_{SQM}$ arcsec$^{-2}$ (Full Moon at 30$^{\circ}$\,from zenith), to 21.3\,mag$_{SQM}$ arcsec$^{-2}$ 
(Galactic plane at the zenith), to the darkest value of 22\,mag$_{SQM}$ arcsec$^{-2}$ on a moonless night. 
A rough estimate of the percentage of the night time (moonless and moonlit nights) free of clouds is found to 
be 74\%, in  accord with previous studies \citep[e.g.,][]{2007RMxAC..31...47T}.\\

\begin{table*}
  \begin{center}
  \begin{threeparttable}
  \caption{Mean night sky brightness at OAN-SPM from 2013 to 2016.} \label{tab:nsb}
 \begin{tabular}{lccccc}
   \hline\hline
\vspace{0.05pt}\\
    Filter & NSB$\pm\sigma$   & NSB$_{min}$  & NSB$_{max}$ & NSB$_{ZL}\pm\sigma$  & $\Delta$ZL \\
           & (mag arcsec$^{-2}$)   & (mag arcsec$^{-2}$)  & (mag arcsec$^{-2}$) & (mag arcsec$^{-2}$) & (mag arcsec$^{-2}$) \\

 \hline
\vspace{0.05pt}\\
    $U$   & 22.27$\pm$0.21  & 21.56  & 22.84 &   22.68$\pm$0.20  &  0.41 \\
    $B$   & 22.60$\pm$0.15  & 22.05  & 23.26 &   23.10$\pm$0.12  &  0.50 \\
    $V$   & 21.59$\pm$0.12  & 21.05  & 22.11 &   21.84$\pm$0.11  &  0.25 \\
    $R$   & 20.90$\pm$0.12  & 20.36  & 21.46 &   21.04$\pm$0.12  &  0.14 \\
    $I$   & 19.32$\pm$0.17  & 18.81  & 19.90 &   19.36$\pm$0.16  &  0.04 \\
    $SQM$ & 21.62$\pm$0.16  & 21.10  & 22.04 &   21.88$\pm$0.15  &  0.26 \\
    \hline
  \end{tabular}
\begin{tablenotes}
      \small
      \item Mean, minimum, and maximum NSB values not corrected for zodiacal light, the mean NSB corrected for 
    zodiacal light, and the mean contribution of the zodiacal light for each filter.  
    The $\sigma$ is the estimated internal error of a individual measurement by subtracting off the yearly averages 
    from the data and computing the Gaussian standard deviation of the resultant distribution.
   \end{tablenotes} 
  \end{threeparttable}
  \end{center}   
\end{table*}

\begin{table}
    \begin{center}
  \begin{threeparttable}
  \caption{Sky colors at OAN-SPM} \label{tab:color}
   \begin{tabular}{lrr}
   \hline\hline
   \vspace{0.05pt}\\
          & Moonless         &  Moonlit\tnote{a}  \\
    Color & Mean $\pm\, \sigma$  & Mean $\pm\, \sigma$ \\
    \hline
    \vspace{0.05pt}\\
    $U-B$ & $-0.323\pm0.174$  & $-0.268\pm0.140$  \\
    $B-V$ &  $1.008\pm0.112$  &  $0.220\pm0.099$   \\
    $V-R$ &  $0.612\pm0.167$  & $-0.003\pm0.024$   \\
    $V-I$ &  $1.521\pm0.370$  &  $0.415\pm0.194$   \\
   \hline
  \end{tabular}
  \begin{tablenotes}
      \small
      \item The $\sigma$ values are calculated as in Table~\ref{tab:nsb}.
      \item[a] Mean values for nine moonlit nights with Moon phase $> 0.75$ and zenithal distances of 40$^{\circ}$--60$^{\circ}$ 
      (Sect.~\ref{sec_moon_present}).
  \end{tablenotes}
  \end{threeparttable}
  \end{center}
\end{table}

We filter the data in several ways in order to minimize unnecessary light contributions from different sources.
First, we only consider data obtained on nights that are entirely clear. This data has been chosen  
by visual inspection of All--sky camera images (see Fig.~\ref{fig:mw_zl}), resulting in 332 full nights.  Second, we 
select only data taken during dark time, when the Sun and Moon are at least 18$^{\circ}$\,below the horizon. Third, 
to minimize the corrections for zodiacal light, we restrict our observations to high helio-ecliptic longitudes. 
Since our data set lies at ecliptic latitudes $\le$55$^{\circ}$, we have applied a correction to all SQM measurements  
depending on their helio-ecliptic coordinates (see Col. 7 in Table~\ref{tab:sqm}) based upon the flux of the zodiacal 
light in the V band (see Fig.~\ref{fig:Zodiacallight_VbandS10}). The contribution of zodiacal light is almost constant 
for a given value of $|\, \beta\, |$ for $|\,\,\lambda-\lambda_{\odot}\,\,|>$110$^{\circ}$, but it varies from 0.2 to 
0.4\,mag from $|\, \beta\, | =  0^{\circ}$\,to $|\, \beta\, | =  55^{\circ}$.  Finally, to avoid strong contributions 
from the Galactic plane, we have restricted our observations to galactic latitudes $|\,b\,|>$20$^{\circ}$. As a 
result, we retain data from 183 nights with at least 10 measurements each. \\ 

Table \ref{tab:sqm} presents all our SQM measurements as well as the zodiacal light corrections we use for each measurement. \\

\section{Results}
\label{sec:results}

\subsection{CCD results}
\label{sec:ccdresults}

In Table~\ref{tab:nsb}, we present the mean, minimum, and maximum values of the NSB before correction for zodiacal 
light (columns 2-4), while in the last two columns we present the values of the NSB corrected for zodiacal light (column 5) and the mean value of 
the zodiacal light contribution in each band ($\Delta$ZL, column 6).  One can see from Col. 6 of Table~\ref{tab:nsb} that the mean 
$\Delta$ZL is as large as 0.5\,mag in the B passband, while for the other filters this correction is smaller. For 
detailed measurements of each eighteen nights and the zodiacal light correction in each filter and 
observing run we refer the reader to Table~\ref{tab:allnights} and Fig~\ref{fig:brillo}. As a by product of 
our CCD sky images we have also determined the sky colors, which we present in Table~\ref{tab:color}. The NSB and 
sky color uncertainties are the estimated internal error of a individual measurement by subtracting off the 
yearly averages from the data and computing the Gaussian standard deviation of the resultant distribution.\\

For two observing runs with Moon phase $> 0.75$ and zenith distance in the range of 40$^{\circ}$\,--60$^{\circ}$, we have 
calculated the NSB and have found that this is on average brighter by 3.3, 3.8, 2.8, 2.1 and 0.7 magnitude in the U, 
B, V, R, and I bands, respectively. In Table~\ref{tab:color} we also give the mean sky colors obtained on 
nine moonlit nights with Moon at zenithal distances in the range of 40$^{\circ}$\,--60$^{\circ}$ and Moon phase $> 0.75$.\\

\subsection{SQM results}
\label{sec:sqmresults}

\begin{figure}[h!]
 \centering
 \includegraphics[width=0.50\textwidth]{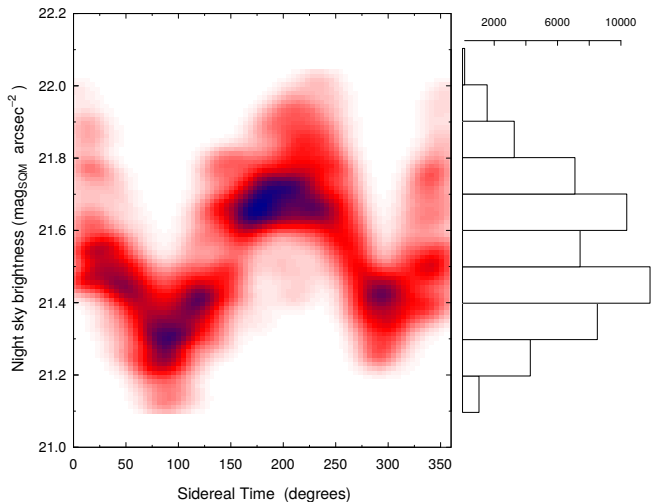}
  \caption{Distribution of the raw SQM measurements as a function of sidereal time (ST) for clear moonless nights. 
  For demonstration purposes, the data have not been filtered to remove observations at low galactic latitude or 
  corrected for the contribution of zodiacal   light.  At the latitude of the OAN-SPM, +31$^{\circ}$, the galactic 
  plane lies at $\mathrm{ST} \sim 90^{\circ}$ and   $\mathrm{ST} \sim 300^{\circ}$. The histogram shows the number of 
  SQM measurements per bin of NSB (bin width is 0.1\,mag  $_{SQM}$ arcsec$^{-2}$). A bimodal distribution is 
  produced due to light from the galactic plane.}\label{fig:NSBtimeST}
\end{figure}

In Fig.~\ref{fig:NSBtimeST} we show the NSB distribution as a function of Sidereal time before filtering for  
Galactic latitude or correcting for the zodiacal light contribution.  At the latitude of OAN-SPM (+31$^{\circ}$), the galactic 
plane lies near Sidereal time 90$^{\circ}$\,and 300$^{\circ}$, where it can be seen that the NSB is brighter (i.e., lower values), 
with a contribution to the sky brightness of about 45\% inside the field of view of the SQM due to the galactic plane.\\

\begin{figure}[h!]
 \centering
 \includegraphics[width=0.50\textwidth]{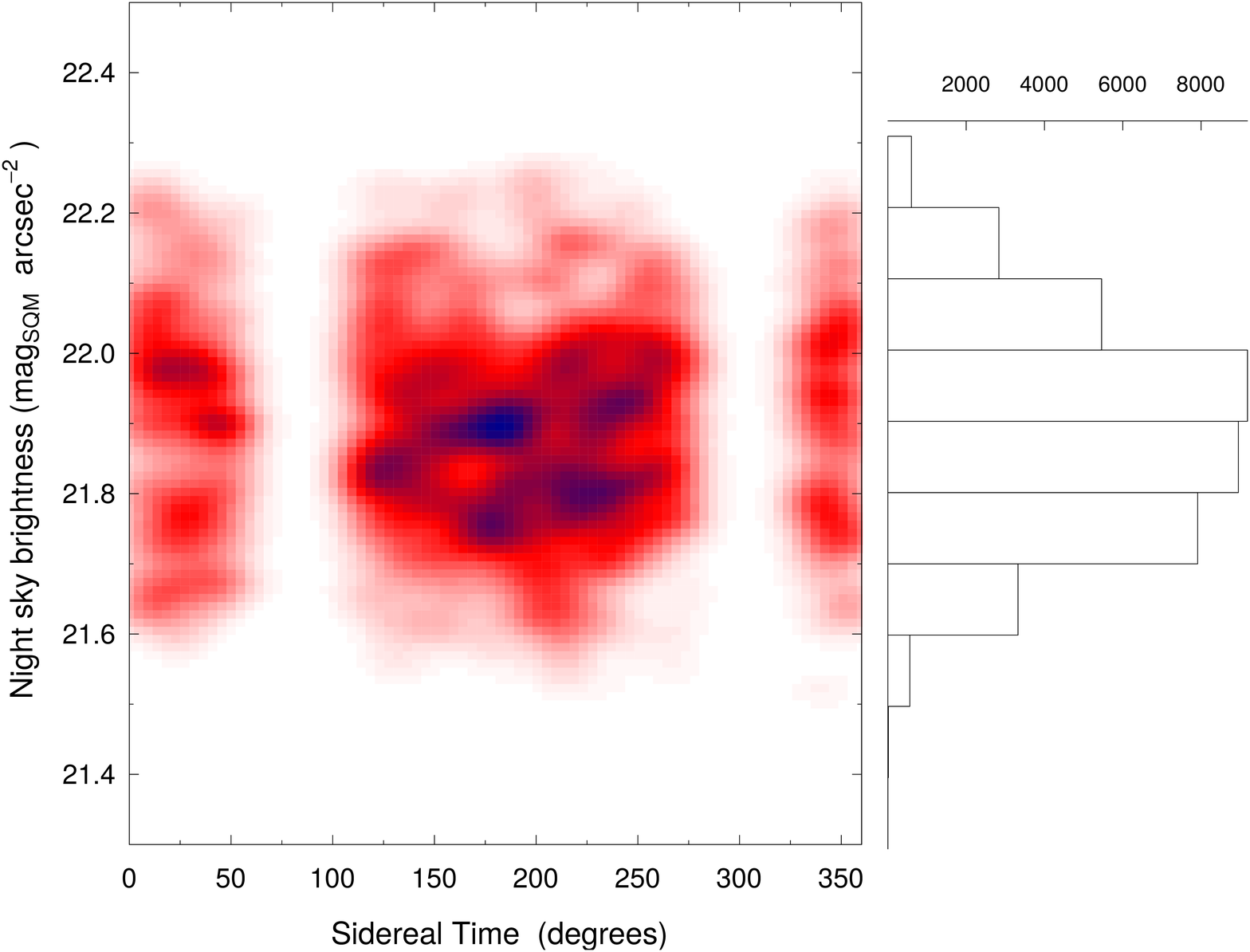}
  \caption{Distribution of the SQM measurements as a function of sidereal time for clear moonless nights. 
  The data is filtered to remove observations at low galactic latitude and corrected to remove the 
  contribution for zodiacal light. The histogram shows the number of SQM measurements per bin of NSB 
  (bin width is 0.1\,mag$_{SQM}$ arcsec$^{-2}$). Compared to Figure \ref{fig:NSBtimeST}, 
  the distribution of NSB values is now narrower and single-peaked.}
  \label{fig:NSBtimeSTZL}
\end{figure}

On the other hand, in Fig.~\ref{fig:NSBtimeSTZL}, we show the SQM data after filtering the data to retain only measurements 
made at high galactic latitude, $|\,b\,| \ge$20$^{\circ}$, and after correcting for the zodiacal light contribution. The 
sky brightness variations are now dominated by the variability in the airglow 
and its patchy structure. The NSB variations in individual nights, from minimum to maximum values, range between 0.1 
and 0.7\,mag$_{SQM}$ arcsec$^{-2}$, with a mean value of 0.2$\pm$0.13\,mag$_{SQM}$ arcsec$^{-2}$. On a given night, the NSB 
may increase, decrease, or remain constant with time. As found by \citet{1988PASP..100..496W}, \citet{1989PASP..101..707P}, 
and \citet{1997PASP..109.1181K}, on any given night the sky brightness can vary 10\% to 50\%. The average dispersion of the NSB on 
a given night in our data is $\pm$0.06\,mag$_{SQM}$ arcsec$^{-2}$ (Table \ref{tab:sqm}, column 3).  \\

\begin{figure}[h!]
 \centering
 \includegraphics[width=0.50\textwidth]{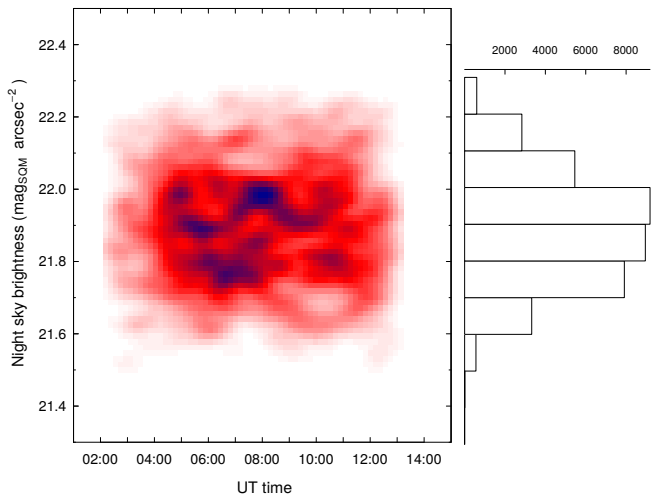}
  \caption{Distribution of the SQM measurements as a function of UT for clear moonless nights. The data is filtered 
  to remove observations at low galactic latitude and corrected to remove the contribution for zodiacal light. 
  The histogram shows the number of SQM measurements per bin of NSB (bin width is 0.1\,mag$_{SQM}$ arcsec$^{-2}$).}
  \label{fig:NSBtimeZL}
\end{figure}

The literature is mixed on whether and how the NSB varies as a function of the time after twilight. \citet{1988PASP..100..496W} 
pointed out that the sky at zenith gets darker by $\sim$0.4\,mag arcsec$^{-2}$ during the first six hours after the end of twilight. 
On the other hand, \citet{1990PASP..102.1052K} found that his data obtained in the V passband showed a decrease of $\sim$0.3\,mag 
arcsec$^{-2}$, but that this was not seen in the B passband. Other authors have not found evidence that the NSB decreases after 
twilight (\citealp{1995A&AS..112...99L}; \citealp{1996A&AS..119..153M}; \citealp{1998NewAR..42..503B}; \citealp{2003A&A...400.1183P}). 
Our data agree with these last results.  In Fig.~\ref{fig:NSBtimeZL}, we present the data set from Fig.~\ref{fig:NSBtimeSTZL}, 
where we now plot the NSB as a function of UT time. From Fig.~\ref{fig:NSBtimeZL}, it can be seen that there is no clear trend 
in NSB after twilight (local midnight at 08:00hrs UT), as found by the work previously cited.  \\

\begin{figure}[h!]
 \centering
 \includegraphics[width=8cm,bb=0 0 1080 2088]{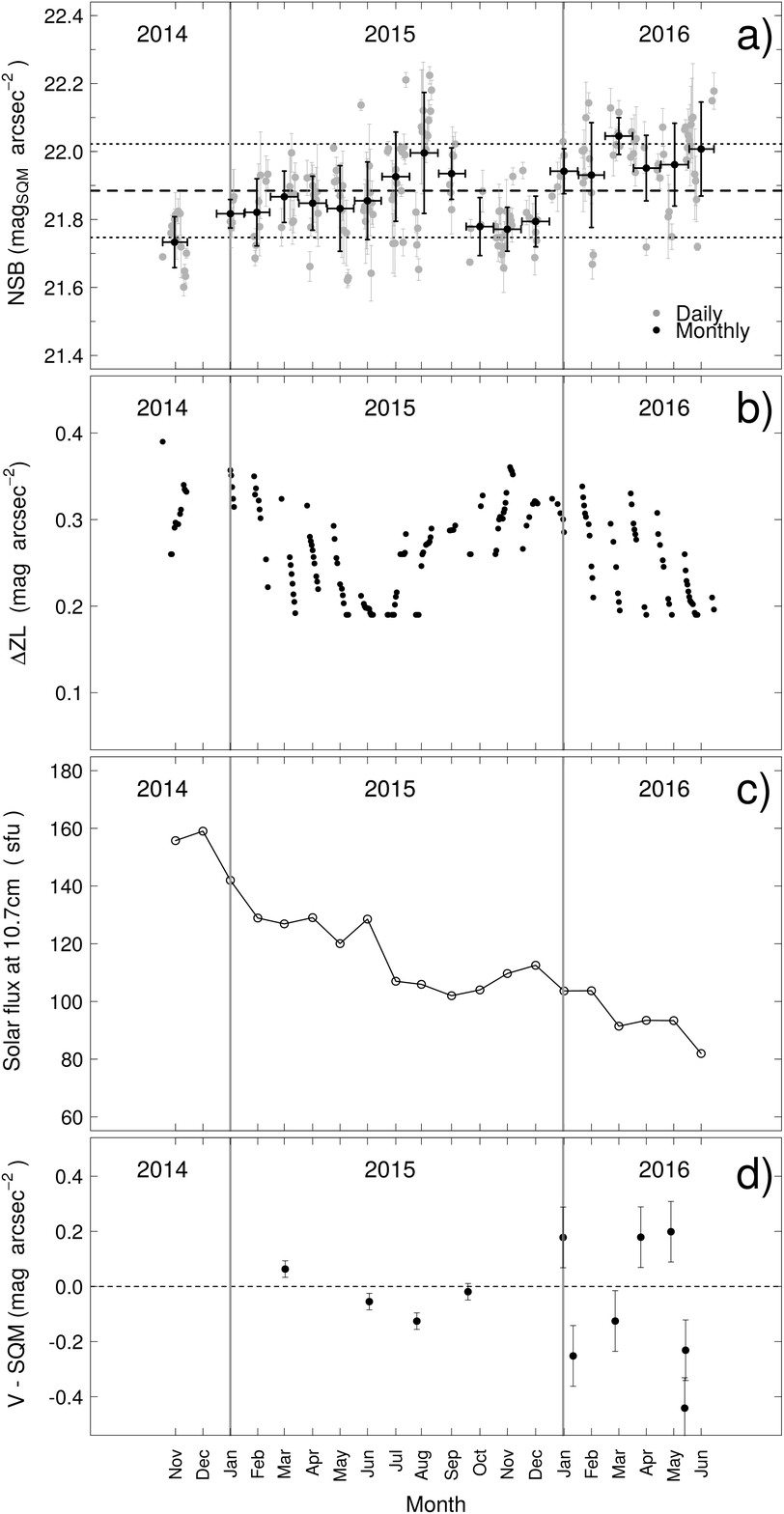}
  \caption{{\bf a)} We present the SQM NSB measured at OAN-SPM from November 2014 to June 2016. The data were obtained on moonless 
  nights, correcting for zodiacal light contributions, and excluding measurements near galactic plane, $|\,b\,| \ge$20$^{\circ}$. 
  The average daily (gray dots) and monthly (black dots) NSB values are shown.  The global average value (dashed line; see 
  last row Table~\ref{tab:nsb}) and the $\pm$1$\sigma$ interval (dotted lines) are plotted.  For the monthly NSB values, 
  the vertical error bars are the 
  standard deviation of the mean values, while horizontal error bars indicate the extent of one month.  No significant seasonal 
  behavior is found in this period. {\bf b)} The zodiacal light contribution applied to NSB measurements. {\bf c)} The monthly solar 
  flux variation. {\bf d)} We plot the difference between CCD V-band and SQM measurements as a function of time.  We use the monthly 
  mean value of $\mathrm{mag}_{SQM}$, since we do not always have a nightly measurement for each date with UBVRI data.  
  In all panels the vertical lines indicate the start and end of the year 2015.}
  \label{fig:NSB_meanZL}
\end{figure}

Besides the variations of the NSB during a single night, there are night-to night and longer-term variations.  
In Fig.~\ref{fig:NSB_meanZL}a, we present daily and monthly mean NSB, which show how the NSB can vary on short 
periods and to investigate any seasonal variation of the NSB.  Although there is considerable variation, 
Fig.~\ref{fig:NSB_meanZL}a shows that there is no evidence for any seasonal variation of the NSB, in the sense 
of a periodic variation.  Instead, all the mean monthly values (black dots) are consistent with the global average given 
in Table~\ref{tab:nsb} (last row) and shown with horizontal dashed lines in Fig.~\ref{fig:NSB_meanZL}a.  Fig.~\ref{fig:NSB_meanZL}b 
presents the zodiacal light correction applied to the data in Fig.~\ref{fig:NSB_meanZL}a (see Table \ref{tab:sqm}).  Meanwhile, 
in Fig.~\ref{fig:NSB_meanZL}c, we plot the variation of the solar flux in the same period, which has been decreasing since 2014.  
The solar flux is expected to be correlated  with the NSB (see Sect.~\ref{sec:solar}).  In Fig.~\ref{fig:NSB_meanZL}d we show the 
differences between CCD V band and SQM measurements in order to check for any drift in the SQM sensor. From this figure it can be seen 
that SQM and CCD measurements are comparable, with no evidence of any drift.\\

\begin{table*}
  \begin{center}
  \begin{threeparttable}
  \caption{Yearly Night sky brightness at OAN-SPM} \label{tab:sun}
\begin{tabular}{lcccc}
\hline\hline    
    \vspace{0.05pt}\\

           & \multicolumn{4}{c}{Year\tnote{a}}      \\
           & \multicolumn{1}{c}{2013}   &   \multicolumn{1}{c}{2014}  &   \multicolumn{1}{c}{2015}  &   \multicolumn{1}{c}{2016}    \\
    Filter & \multicolumn{1}{c}{NSB\,$\pm\,\sigma$} & \multicolumn{1}{c}{NSB\,$\pm\,\sigma$} & \multicolumn{1}{c}{NSB\,$\pm\,\sigma$} & \multicolumn{1}{c}{NSB\,$\pm\,\sigma$}    \\
           & (mag arcsec$^{-2}$) & (mag arcsec$^{-2}$) & (mag arcsec$^{-2}$) & (mag arcsec$^{-2}$) \\
    \hline 
        \vspace{0.05pt}\\
    $U$ &  $22.21\pm0.25$  &  $21.96\pm0.16$   &  $22.29\pm0.16$   &  $22.51\pm0.16$  \\
    $B$ &  $22.59\pm0.22$  &  $22.39\pm0.11$   &  $22.61\pm0.09$   &  $22.74\pm0.17$  \\
    $V$ &  $21.73\pm0.38$  &  $21.45\pm0.17$   &  $21.61\pm0.02$   &  $21.64\pm0.12$  \\
    $R$ &  $20.86\pm0.31$  &  $20.67\pm0.15$   &  $21.06\pm0.04$   &  $20.98\pm0.19$  \\
    $I$ &  $19.38\pm0.50$  &  $19.29\pm0.13$   &  $19.44\pm0.10$   &  $19.27\pm0.24$  \\
  $SQM$ &     ...          &  $21.42\pm0.10$   &  $21.60\pm0.14$   &  $21.72\pm0.12$  \\
    Flux$_\odot$\tnote{b} &  $115$ & $152$  &  $116$  &  $96$  \\
    \hline
  \end{tabular}
  \begin{tablenotes}
      \small
      \item[a] The $\sigma$ is estimated as in Table~\ref{tab:nsb}. NSB values are not 
   corrected for zodiacal light.
      \item[b] Solar 10.7\,cm flux units 1 sfu = 10$^{4}$\,Jy\,=\,10$^{-22}$\,W\, m$^{-2}$\, Hz$^{-1}$.
  \end{tablenotes}
  \end{threeparttable}
  \end{center}
\end{table*}

\subsection{Variation of the NSB with Moon phase and distance}\label{sec_moon_present}

\begin{figure}[h!]
 \centering
 \includegraphics[width=0.50\textwidth]{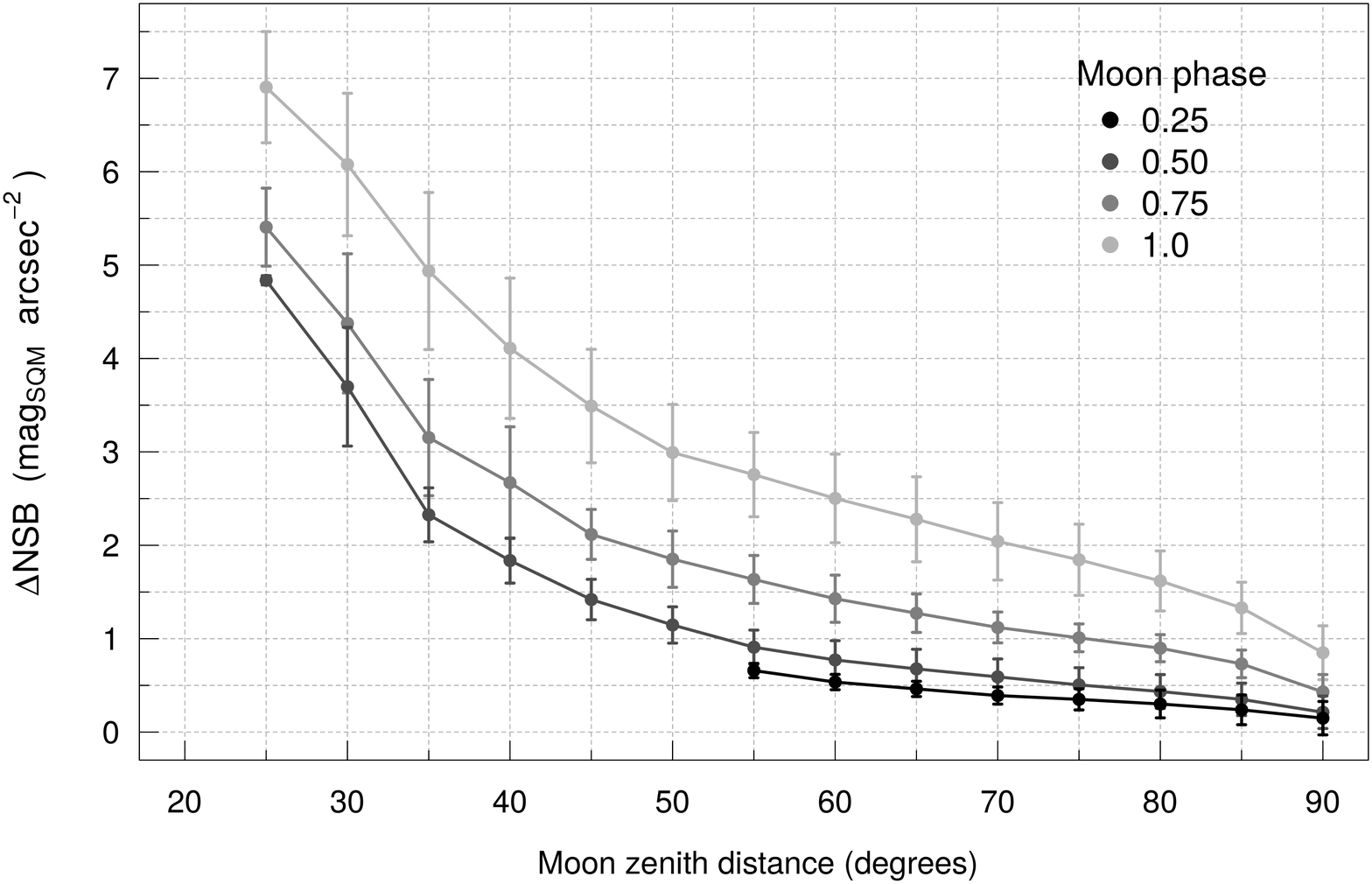}
  \caption{ We plot the mean NSB measured with the SQM as a function of Moon phase and zenith distance.  The error bars represent 
  the standard deviation of the measurements. The NSB at zenith depends sensitively on both Moon phase and zenith distance.}
  \label{fig:Moon_sepNSB}
\end{figure}

As a by product of our SQM  sky  brightness measurements, we performed a quantitative analysis of the data when the 
Moon is above the horizon.  (For this, we use the data filtered from Table \ref{tab:sqm}.)  In Fig.~\ref{fig:Moon_sepNSB}, 
we show the variation in the NSB as a function of Moon phase and zenithal distance.\\

As expected, we find a large increase in the NSB as the zenithal distance of the Moon decreases.  This variation is 
naturally most extreme for the phase of full Moon. Figure~\ref{fig:Moon_sepNSB} should be a useful tool for predicting the sky 
brightness enhancement produced by the presence of the Moon at a given phase and angular distance from an observing 
target, at least in the V band. Taking into account the sky colors on moonlit nights presented in Table~\ref{tab:color}, 
one can see that the sky becomes bluer with the presence of the Moon.  Hence, observations performed with filters bluer 
than V filter, will be more affected by the Moon than redder filters.\\

\section{Discussion}
\label{sec:discuss}

\subsection{Correlation of the NSB with solar activity}
\label{sec:solar}

A correlation between the intensity of the [O~{\sc i}] 5577\,\AA\ airglow line with the sunspot number was reported by 
\citet{1928RSPSA.119...11R} and \citet{1935RSPSA.151...22R}. There is now a well-established correlation with solar activity 
for this and other emission lines, like [O~{\sc i}] 5777, 6300, 6364\, \AA, [O~{\sc ii}] 7320, 7330 \AA, Na D 5890, 5896\,\AA, and OH 
(\citealp{1980GeoRL...7..109A}; \citealp{1981JGR....86.1564Y}; \citealp{1984P&SS...32..897T}). \citet{1988PASP..100..496W} 
also found that there is a correlation between the brightness of the for V and B photometric bands with the solar 10.7\, cm 
radio flux (an indicator of the solar activity), demonstrating that the correlation between sky brightness and solar 
activity applies not only for emission lines, but also for the airglow ``pseudo continuum'' emission. This correlation 
has been confirmed in other studies (\citealp{1989PASP..101..707P}; \citealp{1995A&AS..112...99L}; 
\citealp{1996A&AS..119..153M}; \citealp{1997PASP..109.1181K}).\\

\begin{figure}[h!]
 \centering
 \includegraphics[width=0.50\textwidth]{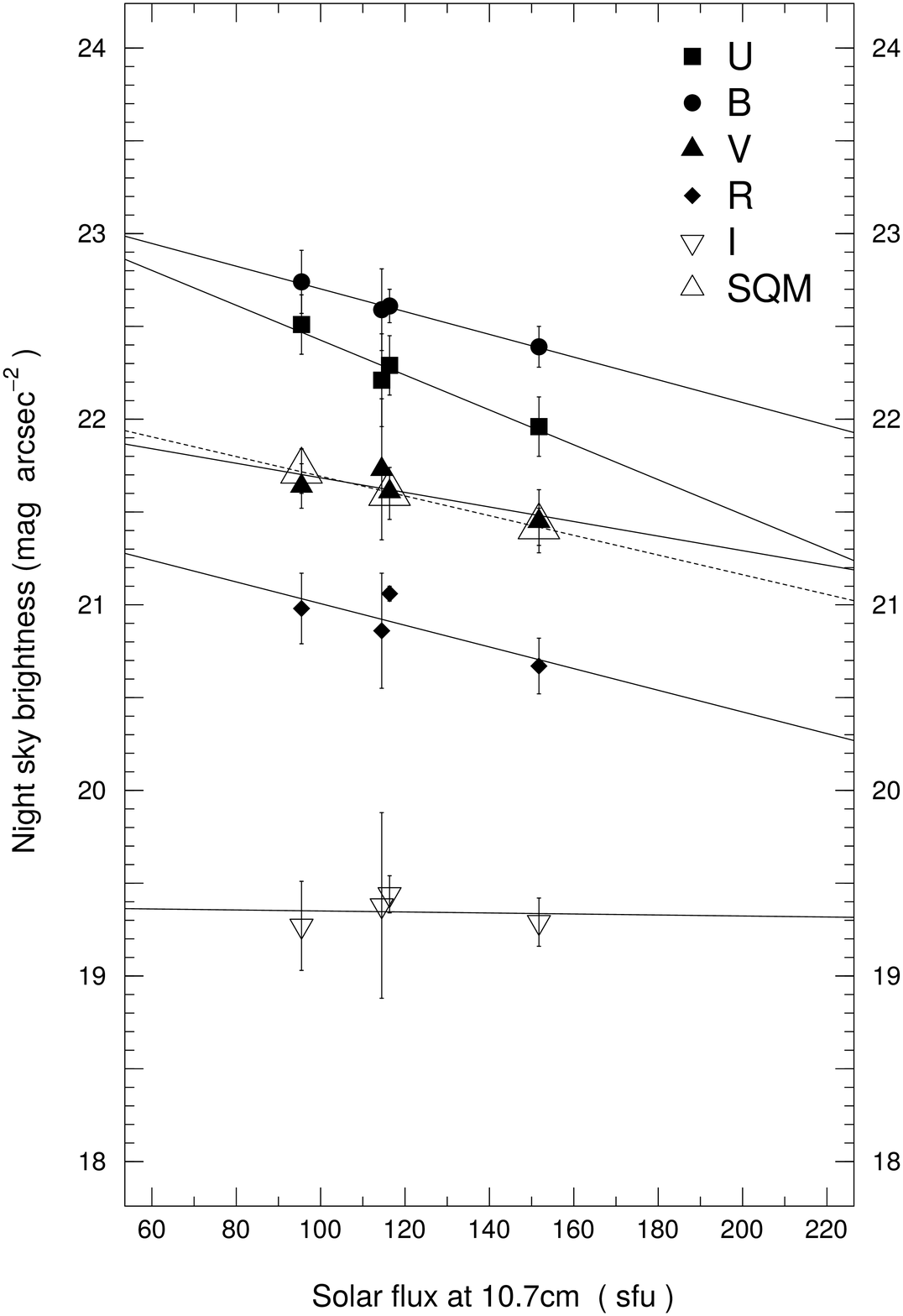}
  \caption{We plot the yearly averages of UBVRI bands and SQM night sky brightness as a function of time (2013-2016). 
  The abscissa values are the averages of the solar 10.7\,cm flux 
  for those months when the NSB was measured. Lines represent fits to each band (solid line for CCD data and dotted 
  line for SQM data).  The detailed fit parameters are presented in Table~\ref{tab:corr}.}
  \label{fig:NSBSolarFluxYearly}
\end{figure}

In order to study this correlation with solar activity in our data, Table \ref{tab:sun} presents the mean NSB for each 
year of our study.  The last row of the table presents the solar 10.7\,cm flux values ``observed'', i.e., not corrected 
to 1 AU solar distance, and averaged over the months in which our NSB observations were made.  The solar fluxes are public 
and provided by the Natural Resources Canada\footnote{\url{http://www.spaceweather.gc.ca}}.  In Fig.~\ref{fig:NSBSolarFluxYearly}, 
we plot the yearly average NSB values for the UBVRI filters and the SQM sensor for 2013--2016 against the solar 10.7\,cm flux (solar 
flux unit; 1 sfu =\,10$^{-22}$\,W\, m$^{-2}$\, Hz$^{-1}$).  We also shown the linear least squares fit to CCD data in each 
filter (solid lines) and SQM data (dotted line).  Our data are from cycle \#24 of the Sun, whose maximum was in early 2014. 
The Sun's activity has been decreasing since then, so we might expect the sky to be darker for the next five years as the Sun 
passes through the minimum in the current cycle. \\

\begin{table}
  \begin{center}
  \begin{threeparttable}
  \caption{Correlations of NSB with solar activity}\label{tab:corr}
 \begin{tabular}{lcccc}
    \hline\hline
        \vspace{0.05pt}\\
    Filter     & a    &  b  & r  & P  \\
    \hline
    \vspace{0.05pt}\\
    $U$  & $-0.009$  & $23.37$  & $-0.97$  &  $0.03$  \\
    $B$  & $-0.006$  & $23.31$  & $-0.99$  &  $0.01$  \\
    $V$  & $-0.004$  & $22.08$  & $-0.79$  &  $0.21$  \\
    $R$  & $-0.006$  & $21.59$  & $-0.81$  &  $0.19$  \\
    $I$  & $-0.0003$ & $19.38$  & $-0.08$  &  $0.92$  \\
  $SQM$  & $-0.005$  & $22.22$  & $-0.99$  &  $0.02$  \\
    \hline
  \end{tabular}
    \begin{tablenotes}
      \small
      \item a and b are the slope and intercept, respectively, of the linear fit, 
        $\mathrm{NSB} = \mathrm a\times \mathrm{Flux}_\odot + \mathrm b$, r is the correlation coefficient, 
        and P is the probability of obtaining the result by chance.
    \end{tablenotes}
  \end{threeparttable}
  \end{center}
\end{table}

In Table~\ref{tab:corr} we present the parameters for the least squares fits shown in Fig.~\ref{fig:NSBSolarFluxYearly}, 
with values of the NSB in each filter and the solar 10.7\,cm flux, $\mathrm{Flux}_\odot$, taken from Table \ref{tab:sun}. 
The correlations in Table~\ref{tab:corr} 
indicate that there is a trend in which the NSB in the UBVR and SQM bands ($|\mathrm r|>0.7$) decreases as the  
solar activity decreases (i.e., lower solar flux), as has been found previously.  These trends have a high statistical 
significance ($\mathrm P < 0.05$) only for the UB and SQM bands.  The U and B bands are nearly devoid of emission lines 
from the sky.  That the significance is lower for the VR bands compared to the SQM-band data is likely the result of 
the much larger number of data averaged in the latter case, given that the slopes are similar in the three cases.  \\

Since 1947, the minimum and maximum of the monthly average solar 10.7\,cm flux are approximately 60 and 250 sfu\footnote{Monthly 
averages from 1947 to 2016 reported by the Natural Resources Canada.}.  If we consider these extreme values and the slopes 
reported in Table~\ref{tab:corr}, the total variation in NSB due to the solar activity over a complete cycle would be 1.8, 1.2, 0.7, 
and 1.1 mag in the UBVR bands, respectively, and 0.9 mag for SQM band. The current solar cycle has been less active, 
varying from a maximum of approximately 162 sfu in early 2014 to 82 sfu in recent months.  This implies that, since early 2014, 
the NSB at the OAN-SPM has decreased by $\Delta U = 0.7$, $\Delta B = 0.5$, $\Delta V = 0.3$, $\Delta R = 0.5$ and 
$\Delta \mathrm{mag}_{SQM} = 0.4$ mag arcsec$^{-2}$, respectively, based upon the slopes in Table \ref{tab:corr}.  \\

\subsection{Comparison with other observing sites}
\label{sec:sites}

\begin{figure}[h!]
 \centering
 \includegraphics[width=0.50\textwidth]{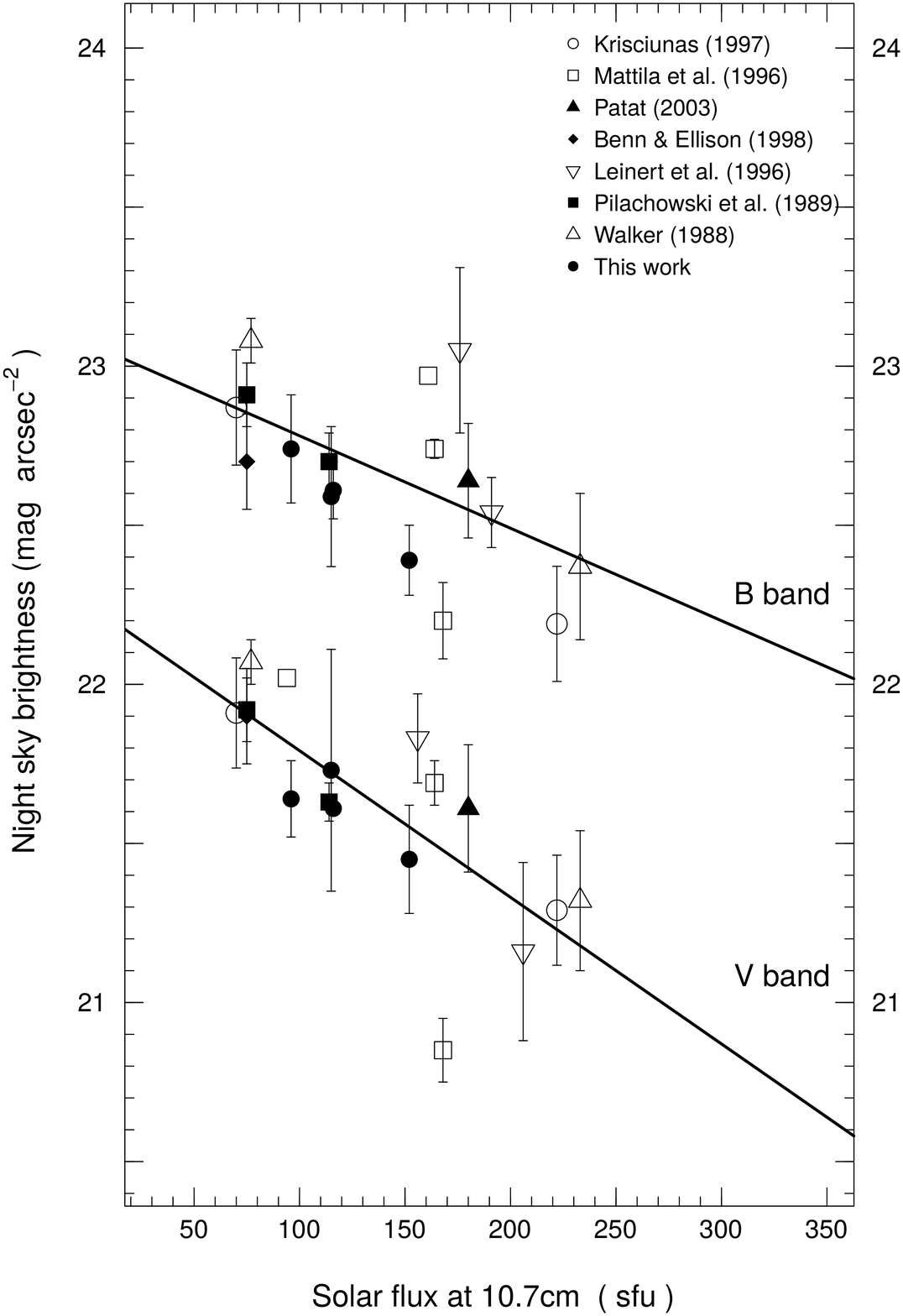}
  \caption{We compare the NSB at different observing sites as a function of the solar activity. The correlation seen here is 
  very similar to that in Figure \ref{fig:NSBSolarFluxYearly}.}
  \label{fig:NSBSolarFluxYearly_comparison}
\end{figure}

When comparing our measurements with the NSB from other observatories, we must account for the solar activity at 
the time all of the measurements were made.  In Table~\ref{tab:sites}, we compare our minimum and maximum 
NSB obtained in 2014 and 2016, respectively (see Table~\ref{tab:sun}), at the OAN-SPM with broad band UBVRI values 
available in the literature for other observatories. In Fig.~\ref{fig:NSBSolarFluxYearly_comparison}, 
we plot the data for the BV bandpasses from Table~\ref{tab:sites}.  This presentation shows that the NSB measured 
at all of these sites are similar once account is taken of the solar activity. \\

In Fig.~\ref{fig:NSBSolarFluxYearly_comparison}, we also show a linear fit to the data in order to determine 
the expected variation of the NSB due to solar activity.  The fit parameters, correlation coefficients, and 
probability that they arise by chance are presented in Table~\ref{tab:corrSites}.  Although there is a lot of 
scatter, the large number of points lead to a robust fit. The dispersion about the fits in 
Table~\ref{tab:corrSites} is not surprising, as it is similar to that found 
in the monthly time bins in our SQM data (Fig.~\ref{fig:NSB_meanZL} and Table~\ref{tab:sqm}). 
From Table \ref{tab:corrSites}, we estimate a maximum variation over a solar cycle of 0.6\,mag and 0.9\,mag in 
the B and V bands, respectively, supposing that the solar activity varies from a minimum of 60 sfu to a maximum of 250 sfu. \\

\begin{table*}
  \begin{center}
  \begin{threeparttable}
  \caption{Comparison of zenith sky brightness at different sites. } \label{tab:sites}
  \tabletypesize{\small}
 \begin{tabular}{lccccccc}
   \hline\hline
       \vspace{0.05pt}\\
   Site      & \multicolumn{1}{c}{NSB$_{U}$} & \multicolumn{1}{c}{NSB$_{B}$}   & \multicolumn{1}{c}{NSB$_{V}$} & \multicolumn{1}{c}{NSB$_{R}$}  & \multicolumn{1}{c}{NSB$_{I}$} & \multicolumn{1}{c}{Flux$_\odot$\tnote{a}} & Ref. \\
             &    &    &   &   &  & (sfu) \\
    \hline
        \vspace{0.05pt}\\
    OAN-SPM          & $21.96$    & $22.39$  &  $21.45$   & $20.67$ & $19.29$  &  $152$ & 1  \\
                     & $22.51$    & $22.70$  &  $21.64$   & $20.98$ & $19.27$  &  $ 96$ &    \\
    \hline
    \vspace{0.05pt}\\
    Hawaii           & ...         & $22.19$  &  $21.29$   & ...      & ...       &  $222$ & 2  \\
                     & ...         & $22.87$  &  $21.91$   & ...      & ...       &  $ 70$ &    \\
    \hline                                                                     
    \vspace{0.05pt}\\
    La Silla         & ...         & $22.20$  &  $20.85$   & ...      & ...       & $168$  & 3  \\
                     & ...         & $22.72$  &  $21.69$   & ...      & ...       & $164$  &    \\
                     & ...         & $22.97$  &  ...        & ...      & ...       & $161$  &    \\
                     & ...         & ...       &  $22.02$   & ...      & ...       & $94$   &    \\
    \hline
    \vspace{0.05pt}\\
    Paranal          & $22.28$    & $22.64$  &  $21.61$   & $20.87$ & $19.71$  & $180$  & 4  \\
    \hline
    \vspace{0.05pt}\\
    La Palma         & $22.00$    & $22.70$  &  $21.90$   & $21.00$ & $20.00$  & $ 75$  & 5  \\
    \hline
    \vspace{0.05pt}\\
    Calar Alto       & ...         & ...       &  $21.16$   & ...      & ...       &  $206$ & 6  \\
                     & ...         & $22.54$  &  ...        & ...      & ...       &  $191$ &    \\
                     & ...         & $23.05$  &  ...        & ...      & ...       &  $176$ &    \\
                     & ...         & ...       &  $21.83$   & ...      & ...       &  $156$ &    \\
    \hline                                                                    
    \vspace{0.05pt}\\
    Kitt Peak        & ...         & $22.70$  &  $21.63$   & ...      & ...       &  $114$ & 7  \\
                     & ...         & $22.91$  &  $21.92$   & ...      & ...       &  $ 75$ &    \\
    \hline                                                                    
    \vspace{0.05pt}\\
    San Benito Mt.   & ...         & $22.37$  &  $21.32$   & ...      & ...       &  $233$ & 8  \\
                     & ...         & $23.08$  &  $22.07$   & ...      & ...       &  $ 77$ &    \\

    \hline
 \end{tabular}
     \begin{tablenotes}
      \small
      \item NSB values are not corrected for zodiacal light contribution. 
      \item[a] Solar 10.7\,cm flux units 1 sfu = 10$^{4}$\,Jy\,=\,10$^{-22}$\,W\, m$^{-2}$\, Hz$^{-1}$.
       \item(1) This work; (2) \cite{1997PASP..109.1181K}; (3) \cite{1996A&AS..119..153M}; (4) \cite{2003A&A...400.1183P};
    (5) \cite{1998NewAR..42..503B}; (6) \cite{1995A&AS..112...99L}; (7) \cite{1989PASP..101..707P}; (8) \cite{1988PASP..100..496W}.
     \end{tablenotes}
  \end{threeparttable}
  \end{center}
\end{table*}

  \begin{center}
  \begin{threeparttable}
  \caption{Correlations of NSB with solar activity for data in Fig.~\ref{fig:NSBSolarFluxYearly_comparison}}  \label{tab:corrSites}
 \begin{tabular}{lccccc}
    \hline\hline
        \vspace{0.05pt}\\
    Filter     & a    &  b  & r  & P  & $\sigma$\\
    \hline
    \vspace{0.05pt}\\
    $B$  & $-0.003$  & $23.07$  & $-0.57$  &  $0.02$     & $0.27$ \\
    $V$  & $-0.005$  & $22.25$  & $-0.76$  &  $0.0004$ & $0.33$ \\
    \hline
  \end{tabular}
       \begin{tablenotes}
      \small
      \item Columns 2 through 5 present, the slope (a), intercept (b), correlation coefficient (r), 
        and the probability (P) for the least-squares fit of a linear relation for each filter, 
        $\mathrm{NSB}= \mathrm a\times \mathrm{Flux}_\odot + \mathrm b$.  Column 6 presents the standard deviation 
        ($\sigma$) about this fit.
       \end{tablenotes}
  \end{threeparttable}
  \end{center}

\section{Summary and conclusions}
\label{sec:concl}

We obtained UBVRI photometry of the night sky brightness (NSB) during 18 nights from 2013 to 2016 and SQM measurements 
on a daily basis from 2014 to 2016 at the Observatorio Astron\'omico Nacional on the Sierra San Pedro M\'artir (OAN-SPM).  
We have taken into account contributions to the sky brightness due to zodiacal light and have excluded observations at
low galactic latitudes in order to compare our data to those obtained at other sites.  We find no clear trend of the 
NSB as a function of time after twilight.  The dispersion of NSB measurements over the course of a night is typically 
0.2\,mag, based upon our SQM data. \\

We investigate the long term variations of the NSB and its correlation with solar activity.  We find a trend of 
decreasing NSB with decreasing solar activity in the UBVR and SQM bands, though the trend is statistically 
robust only for the UB and SQM bands, perhaps due to too few data points in the VR bands. \\

We compare the NSB at the OAN-SPM with measurements made elsewhere and find that the NSB at the OAN-SPM is comparable 
to that of other observing sites.  When comparing data from different observatories, we find a strong correlation 
between the NSB and the solar flux at the time the measurements were made, which can be useful to estimate the 
expected increase in the NSB due to solar actvity for any site.  The variation in the NSB due to solar activity 
can be as high as 0.6 and 0.9\,mag (B and V bands) from the maximum (250 sfu) to the minimum (60 sfu) of the solar cycle. 
The NSB data presented here should be useful for long-term monitoring of the quality of OAN-SPM site, which remains 
one of the darkest sites in use and for future large telescope facilities.\\

\acknowledgments
This work is based upon observations carried out at the Observatorio Astron\'omico 
Nacional on the Sierra San Pedro M\'artir (OAN-SPM), Baja California, M\'exico. 
We thank the daytime and night support staff at the OAN-SPM for facilitating 
and helping obtain our observations. I. P-F. would like to thank L. Guti\'errez, 
M. N\'u\~nez, F. Murillo, M. Murillo, J.~L. Ochoa, F. Quiros, H. Serrano, C. Tejada, 
D. Clark, L. Fox, T. Verdugo, C. Dur\'an, G. Guisa, E. L\'opez, B. Garc\'ia, B. Mart\'inez, 
F. Guill\'en, G. Melgoza, S. Monrroy, and F. Montalvo.

\appendix

\section{Measurements of CCD Night sky brightness}

\begin{figure}[h!]
 \centering
 \includegraphics[width=10cm,bb=0 0 720 1080]{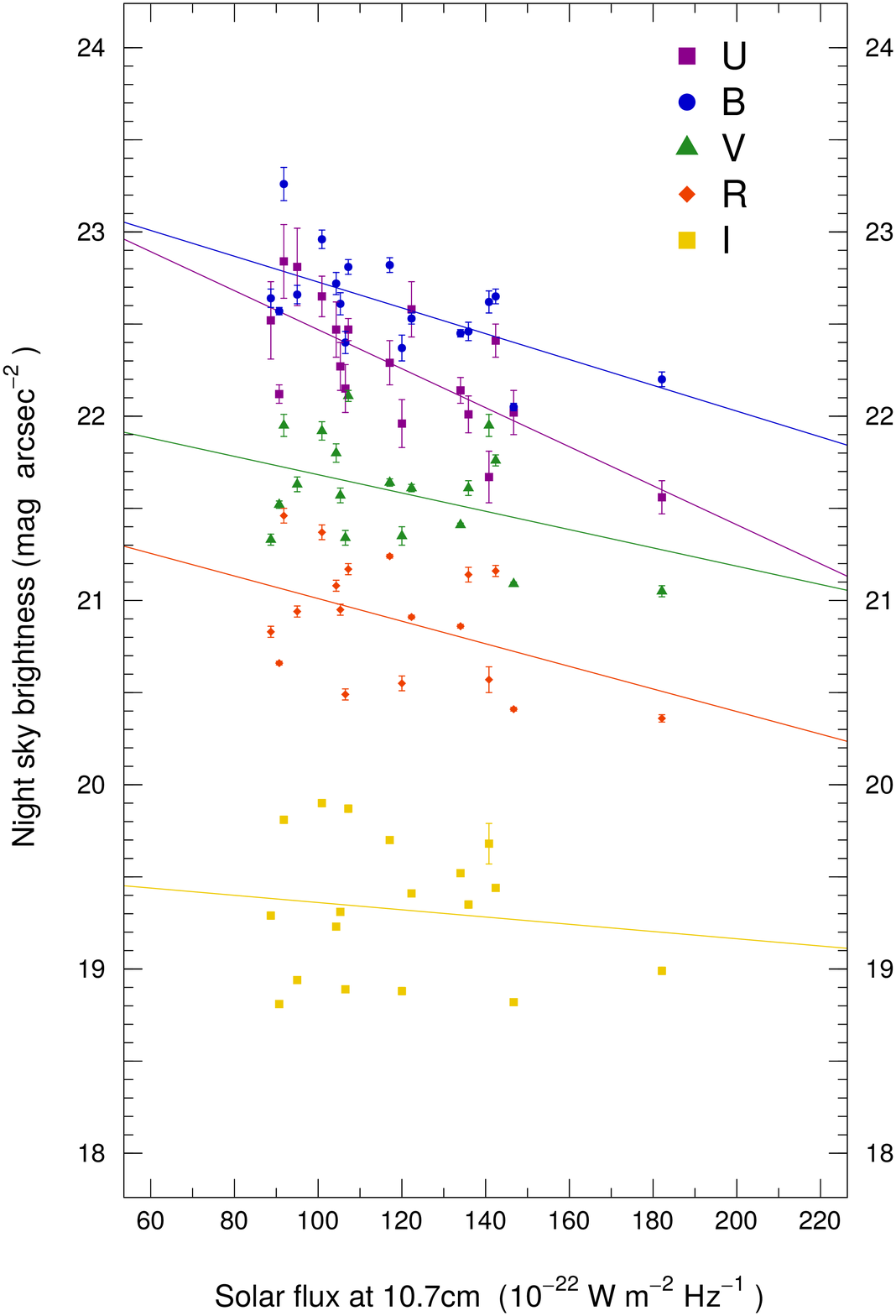}
 \caption{Distribution of the NSB in the UBVRI filters for 18 nights from 2013 to 2016. The abscissa values are 
 the solar 10.7\,cm flux measured on the date of the measurement. The NSB values are not corrected for zodiacal 
 light (see Table~\ref{tab:allnights}).}
  \label{fig:brillo}
\end{figure}

Table \ref{tab:allnights} presents all of the NSB measurements made using CCD images from 18 nights during the years 
2013-2016. Columns 1 and 2 are the UT date and time of the measurements.  Cols. 3-7 present the NSB measurements in 
the UBVRI bands, respectively, uncorrected for the contribution of the zodiacal light.  In parentheses in these columns, 
we present the correction for zodiacal light for each individual measurement.  Column 8 presents the 
``observed'' solar 10.7\,cm flux measured on the previous UT date of the measurement.  Figure \ref{fig:brillo} presents 
these data as a function of the solar 10.7\,cm flux.

\begin{table}
  \begin{center}
  \begin{threeparttable}
\caption{CCD Night sky brightness at OAN-SPM} \label{tab:allnights}
 \begin{tabular}{ccllllll}
   \hline\hline
\vspace{0.05pt}\\
    UT Date  &    UT     & NSB$_{U}$ ($\Delta$ZL)   &  NSB$_{B}$ ($\Delta$ZL)   &  NSB$_{V}$ ($\Delta$ZL)   &  NSB$_{R}$ ($\Delta$ZL)   &   NSB$_{I}$ ($\Delta$ZL) &  \multicolumn{1}{c}{Flux$_\odot$\tnote{a}} \\
             &  (hh:mm)  &              &    &    &   &     \\
 \hline
\vspace{0.05pt}\\
     \multicolumn{2}{c}{2013}           &    &    &   &  &    \\
    18 Feb   &  10:29   &  22.47(0.37)  &  22.81(0.48)   & 22.11(0.24)  &  21.17(0.13)  & 19.87(0.03) &  107 \\ 
     2 Apr   &  04:05   &  21.96(0.50)  &  22.37(0.64)   & 21.35(0.32)  &  20.55(0.18)  & 18.88(0.05) &  120 \\ 
 \hline
\vspace{0.05pt}\\
     \multicolumn{2}{c}{2014}           &    &    &   &  &    \\
    28 Jan   &  11:07   & 22.41(0.38)   &  22.65(0.48)   & 21.76(0.24)  & 21.16(0.13)   & 19.44(0.03) &  142 \\ 
    29 Mar   &  04:45   & 22.02(0.50)   &  22.05(0.61)   & 21.09(0.31)  & 20.41(0.19)   & 18.82(0.05) &  147 \\ 
    24 Apr   &  04:29   & 22.14(0.48)   &  22.45(0.60)   & 21.41(0.31)  & 20.86(0.17)   & 19.52(0.05) &  134 \\ 
    15 Sep   &  05:28   & 21.67(0.33)   &  22.62(0.44)   & 21.95(0.21)  & 20.57(0.11)   & 19.68(0.03) &  140 \\ 
    28 Oct   &  09:11   & 21.56(0.37)   &  22.20(0.47)   & 21.05(0.23)  & 20.36(0.13)   & 18.99(0.03) &  182 \\ 
 \hline
\vspace{0.05pt}\\
     \multicolumn{2}{c}{2015}           &    &    &   &  &    \\
    17 Mar   & 08:10    &  22.29(0.45)  &  22.82(0.55)   & 21.64(0.29)  & 21.24(0.16)   & 19.70(0.04) &  117 \\ 
    17 Jun   & 05:33    &  22.01(0.32)  &  22.46(0.42)   & 21.61(0.19)  & 21.14(0.10)   & 19.35(0.03) &  136 \\ 
     8 Aug   & 06:23    &  22.58(0.53)  &  22.53(0.41)   & 21.61(0.26)  & 20.91(0.14)   & 19.41(0.04) &  122 \\ 
     3 Oct   & 03:22    &  22.27(0.30)  &  22.61(0.40)   & 21.57(0.19)  & 20.95(0.10)   & 19.31(0.03) &  105 \\ 
 \hline
\vspace{0.05pt}\\
     \multicolumn{2}{c}{2016}           &    &    &   &  &    \\
    15 Jan   & 09:24    &  22.47(0.50)  & 22.72(0.64)    & 21.80(0.32)  & 21.08(0.18)   & 19.23(0.05) &  104 \\ 
    26 Jan   & 02:54    &  22.15(0.56)  & 22.40(0.69)    & 21.34(0.35)  & 20.49(0.20)   & 18.89(0.05) &  107 \\ 
    12 Mar   & 08:53    &  22.81(0.45)  & 22.66(0.55)    & 21.63(0.29)  & 20.94(0.16)   & 18.94(0.04) &  95 \\ 
     9 Apr   & 09:39    &  22.65(0.35)  & 22.96(0.46)    & 21.92(0.21)  & 21.37(0.12)   & 19.90(0.03) &  101 \\ 
    12 May   & 11:14    &  22.84(0.32)  & 23.26(0.40)    & 21.95(0.21)  & 21.46(0.10)   & 19.81(0.03) &  92 \\ 
    27 May   & 06:41    &  22.52(0.28)  & 22.64(0.41)    & 21.33(0.19)  & 20.83(0.14)   & 19.29(0.03) &  89 \\ 
    28 May   & 04:57    &  22.12(0.35)  & 22.57(0.45)    & 21.52(0.21)  & 20.66(0.11)   & 18.81(0.03) &  91 \\ 

    \hline

  \end{tabular}
         \begin{tablenotes}
      \small
      \item Units are mag arcsec$^{-2}$. $\Delta$ZL is the correction for zodiacal light. 
      \item Units are 1 sfu = 10$^{4}$\,Jy\,=\,10$^{-22}$\,W\, m$^{-2}$\, Hz$^{-1}$.
      \item[a] Solar flux measurements are the last values reported from the previous UT date.
         \end{tablenotes}
  \end{threeparttable}
  \end{center}
\end{table}

\section{Measurements of SQM Night sky brightness}

Table \ref{tab:sqm} presents all of the NSB measurements made with the SQM sensor for 183 clear nights during dark 
time from November 2014 to June 2016. Column 1 is the UT date, Col. 2 the mean NSB, uncorrected for the contribution 
of the zodiacal light, Col. 3 the standard deviation of the measurements, Cols. 4 and 5 the minimum NSB and maximum NSB, 
respectively, Col. 6 is the number of measurements (one per minute), and Col. 7 is the zodiacal light correction.

\clearpage

\begin{center}
\begin{scriptsize}
\begin{longtable}{lrrrrrr}
\caption[]{SQM Night sky brightness at OAN-SPM}\label{tab:sqm}\\[0.5ex]
\hline\hline \\[2ex]
     UT Date    &  NSB  &  $\sigma$ &  NSB$_{min}$  &  NSB$_{max}$  &  N  &   $\Delta$ZL \\[2ex]
\hline \\[1.8ex]     
\endfirsthead
\caption{continued.}\\
\hline\hline \\[2ex]
     UT Date    &  NSB  &  $\sigma$ &  NSB$_{min}$  &  NSB$_{max}$  &  N  &   $\Delta$ZL \\[2ex]
\hline \\[1.8ex]     
\endhead
\hline
\endfoot
2014-11-03   &    21.30   &    0.000   &    21.30   &    21.30   &    15   &    0.39     \\[0.5ex]
2014-11-12   &    21.50   &    0.022   &    21.46   &    21.54   &    57   &    0.26     \\[0.5ex]
2014-11-13   &    21.52   &    0.031   &    21.45   &    21.57   &    117   &    0.26    \\[0.5ex]
2014-11-16   &    21.51   &    0.037   &    21.40   &    21.55   &    287   &    0.29    \\[0.5ex]
2014-11-17   &    21.52   &    0.032   &    21.46   &    21.59   &    316   &    0.30    \\[0.5ex]
2014-11-18   &    21.45   &    0.043   &    21.33   &    21.50   &    316   &    0.30    \\[0.5ex]
2014-11-20   &    21.53   &    0.037   &    21.43   &    21.57   &    306   &    0.29    \\[0.5ex]
2014-11-22   &    21.51   &    0.100   &    21.36   &    21.68   &    396   &    0.31    \\[0.5ex]
2014-11-23   &    21.41   &    0.126   &    21.23   &    21.61   &    363   &    0.31    \\[0.5ex]
2014-11-26   &    21.26   &    0.017   &    21.23   &    21.30   &    190   &    0.34    \\[0.5ex]
2014-11-27   &    21.31   &    0.014   &    21.25   &    21.33   &    133   &    0.33    \\[0.5ex]
2014-11-28   &    21.30   &    0.010   &    21.27   &    21.32   &    131   &    0.33    \\[0.5ex]
2014-11-29   &    21.37   &    0.019   &    21.34   &    21.40   &    134   &    0.33    \\[0.5ex]
2015-01-16   &    21.44   &    0.024   &    21.40   &    21.48   &    162   &    0.36    \\[0.5ex]
2015-01-17   &    21.42   &    0.063   &    21.36   &    21.53   &    220   &    0.35    \\[0.5ex]
2015-01-18   &    21.45   &    0.064   &    21.36   &    21.58   &    273   &    0.34    \\[0.5ex]
2015-01-19   &    21.53   &    0.133   &    21.37   &    21.77   &    329   &    0.32    \\[0.5ex]
2015-01-20   &    21.55   &    0.118   &    21.40   &    21.75   &    385   &    0.31    \\[0.5ex]
2015-02-11   &    21.40   &    0.000   &    21.40   &    21.40   &    11    &    0.35    \\[0.5ex]
2015-02-12   &    21.36   &    0.008   &    21.34   &    21.38   &    72    &    0.33    \\[0.9ex]
2015-02-13   &    21.39   &    0.040   &    21.34   &    21.46   &    131   &    0.34    \\[0.5ex]
2015-02-16   &    21.53   &    0.121   &    21.36   &    21.70   &    309   &    0.32    \\[0.5ex]
2015-02-17   &    21.47   &    0.102   &    21.32   &    21.62   &    369   &    0.31    \\[0.5ex]
2015-02-18   &    21.63   &    0.153   &    21.39   &    21.81   &    422   &    0.30    \\[0.5ex]
2015-02-24   &    21.66   &    0.043   &    21.59   &    21.73   &    235   &    0.25    \\[0.5ex]
2015-02-26   &    21.71   &    0.027   &    21.67   &    21.74   &    113   &    0.22    \\[0.5ex]
2015-03-13   &    21.45   &    0.054   &    21.39   &    21.61   &    166   &    0.32    \\[0.5ex]
2015-03-22   &    21.64   &    0.090   &    21.42   &    21.79   &    408   &    0.26    \\[0.5ex]
2015-03-23   &    21.63   &    0.082   &    21.46   &    21.74   &    339   &    0.25    \\[0.5ex]
2015-03-24   &    21.76   &    0.043   &    21.64   &    21.82   &    271   &    0.24    \\[0.5ex]
2015-03-25   &    21.57   &    0.041   &    21.50   &    21.64   &    208   &    0.23    \\[0.5ex]
2015-03-26   &    21.58   &    0.020   &    21.55   &    21.62   &    151   &    0.21    \\[0.5ex]
2015-03-27   &    21.67   &    0.052   &    21.55   &    21.73   &    99   &    0.20     \\[0.5ex]
2015-03-28   &    21.73   &    0.039   &    21.67   &    21.79   &    53   &    0.19     \\[0.5ex]
2015-04-10   &    21.61   &    0.083   &    21.48   &    21.73   &    104   &    0.32    \\[0.5ex]
2015-04-13   &    21.38   &    0.055   &    21.26   &    21.50   &    250   &    0.28    \\[0.5ex]
2015-04-14   &    21.50   &    0.051   &    21.33   &    21.58   &    298   &    0.28    \\[0.5ex]
2015-04-15   &    21.66   &    0.066   &    21.46   &    21.72   &    344   &    0.27    \\[0.5ex]
2015-04-16   &    21.63   &    0.084   &    21.40   &    21.75   &    385   &    0.26    \\[0.5ex]
2015-04-17   &    21.62   &    0.050   &    21.43   &    21.69   &    421   &    0.26    \\[0.5ex]
2015-04-18   &    21.61   &    0.063   &    21.43   &    21.70   &    461   &    0.25    \\[0.5ex]
2015-04-20   &    21.62   &    0.073   &    21.36   &    21.68   &    375   &    0.23    \\[0.5ex]
2015-04-21   &    21.65   &    0.114   &    21.35   &    21.76   &    307   &    0.23    \\[0.5ex]
2015-04-22   &    21.60   &    0.072   &    21.42   &    21.69   &    245   &    0.22    \\[0.5ex]
2015-05-09   &    21.72   &    0.024   &    21.68   &    21.75   &    55   &    0.29     \\[0.5ex]
2015-05-10   &    21.63   &    0.040   &    21.57   &    21.68   &    102   &    0.28    \\[0.5ex]
2015-05-12   &    21.69   &    0.051   &    21.62   &    21.79   &    191   &    0.26    \\[0.5ex]
2015-05-13   &    21.67   &    0.062   &    21.55   &    21.75   &    235   &    0.25    \\[0.5ex]
2015-05-16   &    21.64   &    0.067   &    21.45   &    21.74   &    349   &    0.23    \\[0.5ex]
2015-05-18   &    21.61   &    0.120   &    21.34   &    21.74   &    362   &    0.22    \\[0.5ex]
2015-05-19   &    21.69   &    0.084   &    21.43   &    21.79   &    306   &    0.21    \\[0.5ex]
2015-05-20   &    21.56   &    0.086   &    21.35   &    21.68   &    244   &    0.20    \\[0.5ex]
2015-05-23   &    21.57   &    0.096   &    21.39   &    21.71   &    95   &    0.19     \\[0.5ex]
2015-05-24   &    21.43   &    0.022   &    21.38   &    21.45   &    55   &    0.19     \\[0.5ex]
2015-05-25   &    21.44   &    0.024   &    21.40   &    21.48   &    17   &    0.19     \\[0.5ex]
2015-06-08   &    21.92   &    0.017   &    21.89   &    21.95   &    51   &    0.21     \\[0.5ex]
2015-06-11   &    21.62   &    0.023   &    21.57   &    21.65   &    165   &    0.20    \\[0.5ex]
2015-06-12   &    21.64   &    0.060   &    21.49   &    21.72   &    205   &    0.20    \\[0.5ex]
2015-06-13   &    21.60   &    0.116   &    21.29   &    21.72   &    239   &    0.20    \\[0.5ex]
2015-06-14   &    21.66   &    0.135   &    21.36   &    21.81   &    235   &    0.20    \\[0.5ex]
2015-06-15   &    21.64   &    0.132   &    21.43   &    21.85   &    230   &    0.20    \\[0.5ex]
2015-06-16   &    21.63   &    0.082   &    21.42   &    21.77   &    226   &    0.20    \\[0.5ex]
2015-06-17   &    21.76   &    0.116   &    21.52   &    21.92   &    213   &    0.20    \\[0.5ex]
2015-06-18   &    21.69   &    0.125   &    21.43   &    21.84   &    160   &    0.19    \\[0.5ex]
2015-06-19   &    21.45   &    0.082   &    21.33   &    21.58   &    112   &    0.19    \\[0.5ex]
2015-06-20   &    21.66   &    0.064   &    21.55   &    21.77   &    71   &    0.19     \\[0.5ex]
2015-06-21   &    21.62   &    0.038   &    21.56   &    21.69   &    32   &    0.19     \\[0.5ex]
2015-07-07   &    21.81   &    0.007   &    21.80   &    21.82   &    10   &    0.19     \\[0.5ex]
2015-07-08   &    21.82   &    0.021   &    21.77   &    21.84   &    46   &    0.19     \\[0.5ex]
2015-07-12   &    21.67   &    0.115   &    21.45   &    21.86   &    125   &    0.19    \\[0.5ex]
2015-07-13   &    21.54   &    0.087   &    21.38   &    21.65   &    121   &    0.19    \\[0.5ex]
2015-07-14   &    21.68   &    0.091   &    21.52   &    21.80   &    119   &    0.19    \\[0.5ex]
2015-07-15   &    21.53   &    0.100   &    21.36   &    21.69   &    138   &    0.20    \\[0.5ex]
2015-07-16   &    21.74   &    0.077   &    21.57   &    21.83   &    158   &    0.21    \\[0.5ex]
2015-07-17   &    21.69   &    0.077   &    21.52   &    21.78   &    140   &    0.22    \\[0.5ex]
2015-07-21   &    21.75   &    0.014   &    21.72   &    21.77   &    71   &    0.26     \\[0.5ex]
2015-07-22   &    21.62   &    0.017   &    21.57   &    21.64   &    75   &    0.26     \\[0.5ex]
2015-07-23   &    21.74   &    0.018   &    21.70   &    21.76   &    80   &    0.26     \\[0.5ex]
2015-07-24   &    21.47   &    0.040   &    21.41   &    21.52   &    85   &    0.26     \\[0.5ex]
2015-07-25   &    21.75   &    0.038   &    21.69   &    21.83   &    91   &    0.26     \\[0.5ex]
2015-07-26   &    21.73   &    0.024   &    21.69   &    21.75   &    58   &    0.26     \\[0.5ex]
2015-07-27   &    21.93   &    0.006   &    21.92   &    21.94   &    15   &    0.28     \\[0.5ex]
2015-08-07   &    21.63   &    0.058   &    21.55   &    21.74   &    47   &    0.19     \\[0.5ex]
2015-08-08   &    21.54   &    0.034   &    21.48   &    21.59   &    44   &    0.19     \\[0.5ex]
2015-08-09   &    21.58   &    0.040   &    21.52   &    21.64   &    40   &    0.19     \\[0.5ex]
2015-08-10   &    21.46   &    0.032   &    21.41   &    21.51   &    38   &    0.19     \\[0.5ex]
2015-08-13   &    21.83   &    0.144   &    21.55   &    22.02   &    154   &    0.25    \\[0.5ex]
2015-08-14   &    21.80   &    0.154   &    21.40   &    21.92   &    212   &    0.26    \\[0.5ex]
2015-08-15   &    21.86   &    0.124   &    21.52   &    22.00   &    215   &    0.26    \\[0.5ex]
2015-08-18   &    21.73   &    0.051   &    21.64   &    21.81   &    207   &    0.27    \\[0.5ex]
2015-08-19   &    21.78   &    0.028   &    21.68   &    21.82   &    212   &    0.27    \\[0.5ex]
2015-08-20   &    21.77   &    0.051   &    21.65   &    21.85   &    216   &    0.27    \\[0.5ex]
2015-08-21   &    21.82   &    0.047   &    21.72   &    21.87   &    221   &    0.27    \\[0.5ex]
2015-08-22   &    21.95   &    0.039   &    21.85   &    22.00   &    208   &    0.27    \\[0.5ex]
2015-08-23   &    21.84   &    0.028   &    21.76   &    21.88   &    168   &    0.28    \\[0.5ex]
2015-08-24   &    21.89   &    0.017   &    21.84   &    21.91   &    118   &    0.29    \\[0.5ex]
2015-09-13   &    21.61   &    0.047   &    21.50   &    21.69   &    320   &    0.29    \\[0.5ex]
2015-09-14   &    21.59   &    0.037   &    21.52   &    21.65   &    321   &    0.29    \\[0.5ex]
2015-09-15   &    21.54   &    0.051   &    21.43   &    21.64   &    318   &    0.29    \\[0.5ex]
2015-09-16   &    21.70   &    0.079   &    21.10   &    21.79   &    318   &    0.29    \\[0.5ex]
2015-09-17   &    21.70   &    0.049   &    21.58   &    21.76   &    321   &    0.29    \\[0.5ex]
2015-09-19   &    21.73   &    0.035   &    21.64   &    21.79   &    277   &    0.29    \\[0.5ex]
2015-10-05   &    21.41   &    0.009   &    21.40   &    21.43   &    52   &    0.26     \\[0.5ex]
2015-10-06   &    21.51   &    0.026   &    21.48   &    21.56   &    174   &    0.26    \\[0.5ex]
2015-10-17   &    21.47   &    0.024   &    21.42   &    21.50   &    496   &    0.32    \\[0.5ex]
2015-10-19   &    21.56   &    0.049   &    21.46   &    21.63   &    280   &    0.33    \\[0.5ex]
2015-11-02   &    21.52   &    0.015   &    21.50   &    21.56   &    64   &    0.26     \\[0.5ex]
2015-11-03   &    21.48   &    0.013   &    21.46   &    21.51   &    126   &    0.26    \\[0.5ex]
2015-11-05   &    21.43   &    0.020   &    21.37   &    21.47   &    245   &    0.29    \\[0.5ex]
2015-11-06   &    21.52   &    0.044   &    21.41   &    21.58   &    302   &    0.30    \\[0.5ex]
2015-11-07   &    21.52   &    0.044   &    21.40   &    21.57   &    326   &    0.30    \\[0.5ex]
2015-11-08   &    21.44   &    0.058   &    21.35   &    21.54   &    326   &    0.30    \\[0.5ex]
2015-11-09   &    21.39   &    0.052   &    21.29   &    21.47   &    326   &    0.30    \\[0.5ex]
2015-11-10   &    21.42   &    0.037   &    21.34   &    21.48   &    326   &    0.30    \\[0.5ex]
2015-11-11   &    21.35   &    0.096   &    21.11   &    21.48   &    361   &    0.31    \\[0.5ex]
2015-11-12   &    21.41   &    0.076   &    21.19   &    21.47   &    362   &    0.31    \\[0.5ex]
2015-11-13   &    21.44   &    0.083   &    21.26   &    21.55   &    318   &    0.32    \\[0.5ex]
2015-11-14   &    21.49   &    0.121   &    21.30   &    21.67   &    274   &    0.33    \\[0.5ex]
2015-11-18   &    21.45   &    0.036   &    21.39   &    21.50   &    83   &    0.36     \\[0.5ex]
2015-11-19   &    21.44   &    0.010   &    21.42   &    21.45   &    87   &    0.36     \\[0.5ex]
2015-11-20   &    21.43   &    0.058   &    21.32   &    21.51   &    90   &    0.36     \\[0.5ex]
2015-11-21   &    21.57   &    0.020   &    21.53   &    21.61   &    94   &    0.35     \\[0.5ex]
2015-12-02   &    21.68   &    0.023   &    21.63   &    21.71   &    154   &    0.27    \\[0.5ex]
2015-12-06   &    21.53   &    0.042   &    21.46   &    21.61   &    247   &    0.29    \\[0.5ex]
2015-12-09   &    21.50   &    0.084   &    21.37   &    21.64   &    300   &    0.30    \\[0.5ex]
2015-12-13   &    21.48   &    0.056   &    21.40   &    21.60   &    320   &    0.32    \\[0.5ex]
2015-12-15   &    21.37   &    0.049   &    21.31   &    21.45   &    208   &    0.32    \\[0.5ex]
2015-12-16   &    21.44   &    0.047   &    21.36   &    21.50   &    211   &    0.32    \\[0.5ex]
2015-12-17   &    21.42   &    0.022   &    21.37   &    21.47   &    216   &    0.32    \\[0.5ex]
2015-12-18   &    21.48   &    0.026   &    21.41   &    21.52   &    220   &    0.32    \\[0.5ex]
2016-01-03   &    21.54   &    0.014   &    21.52   &    21.57   &    122   &    0.32    \\[0.5ex]
2016-01-09   &    21.58   &    0.082   &    21.44   &    21.69   &    372   &    0.32    \\[0.5ex]
2016-01-12   &    21.62   &    0.038   &    21.52   &    21.68   &    326   &    0.31    \\[0.5ex]
2016-01-15   &    21.73   &    0.065   &    21.52   &    21.81   &    319   &    0.30    \\[0.5ex]
2016-01-16   &    21.70   &    0.037   &    21.65   &    21.77   &    251   &    0.29    \\[0.5ex]
2016-02-05   &    21.66   &    0.026   &    21.59   &    21.73   &    248   &    0.34    \\[0.5ex]
2016-02-06   &    21.58   &    0.108   &    21.37   &    21.74   &    303   &    0.33    \\[0.5ex]
2016-02-07   &    21.69   &    0.082   &    21.50   &    21.82   &    355   &    0.32    \\[0.5ex]
2016-02-08   &    21.62   &    0.099   &    21.46   &    21.86   &    408   &    0.31    \\[0.5ex]
2016-02-09   &    21.80   &    0.142   &    21.52   &    22.00   &    425   &    0.30    \\[0.5ex]
2016-02-12   &    21.85   &    0.061   &    21.77   &    21.95   &    740   &    0.29    \\[0.5ex]
2016-02-13   &    21.70   &    0.052   &    21.64   &    21.82   &    608   &    0.28    \\[0.5ex]
2016-02-15   &    21.63   &    0.010   &    21.61   &    21.66   &    332   &    0.25    \\[0.5ex]
2016-02-16   &    21.44   &    0.028   &    21.40   &    21.50   &    208   &    0.23    \\[0.5ex]
2016-02-17   &    21.49   &    0.014   &    21.46   &    21.50   &    92   &    0.21     \\[0.5ex]
2016-03-07   &    21.83   &    0.084   &    21.61   &    21.95   &    426   &    0.30    \\[0.5ex]
2016-03-10   &    21.71   &    0.037   &    21.64   &    21.82   &    481   &    0.27    \\[0.5ex]
2016-03-13   &    21.77   &    0.029   &    21.71   &    21.82   &    271   &    0.25    \\[0.5ex]
2016-03-15   &    21.84   &    0.021   &    21.81   &    21.87   &    138   &    0.21    \\[0.5ex]
2016-03-16   &    21.91   &    0.027   &    21.85   &    21.95   &    85   &    0.20     \\[0.5ex]
2016-03-17   &    21.82   &    0.019   &    21.79   &    21.85   &    36   &    0.20     \\[0.5ex]
2016-03-29   &    21.65   &    0.039   &    21.58   &    21.70   &    91   &    0.33     \\[0.5ex]
2016-03-30   &    21.74   &    0.026   &    21.69   &    21.79   &    143   &    0.32    \\[0.5ex]
2016-04-01   &    21.70   &    0.045   &    21.61   &    21.79   &    241   &    0.30    \\[0.5ex]
2016-04-02   &    21.74   &    0.098   &    21.57   &    21.90   &    288   &    0.29    \\[0.5ex]
2016-04-03   &    21.66   &    0.057   &    21.55   &    21.76   &    334   &    0.28    \\[0.5ex]
2016-04-04   &    21.76   &    0.112   &    21.62   &    21.93   &    379   &    0.28    \\[0.5ex]
2016-04-13   &    21.81   &    0.066   &    21.64   &    21.88   &    117   &    0.20    \\[0.5ex]
2016-04-15   &    21.53   &    0.025   &    21.49   &    21.57   &    25   &    0.19     \\[0.5ex]
2016-04-27   &    21.64   &    0.027   &    21.59   &    21.69   &    44   &    0.31     \\[0.5ex]
2016-04-28   &    21.68   &    0.049   &    21.57   &    21.74   &    180   &    0.28    \\[0.5ex]
2016-04-30   &    21.65   &    0.046   &    21.55   &    21.72   &    372   &    0.27    \\[0.5ex]
2016-05-03   &    21.82   &    0.073   &    21.61   &    21.87   &    310   &    0.25    \\[0.5ex]
2016-05-04   &    21.75   &    0.031   &    21.64   &    21.79   &    346   &    0.25    \\[0.5ex]
2016-05-09   &    21.60   &    0.107   &    21.35   &    21.72   &    255   &    0.21    \\[0.5ex]
2016-05-10   &    21.62   &    0.075   &    21.42   &    21.70   &    191   &    0.20    \\[0.5ex]
2016-05-13   &    21.56   &    0.037   &    21.50   &    21.61   &    45   &    0.19     \\[0.5ex]
2016-05-27   &    21.80   &    0.008   &    21.79   &    21.82   &    39   &    0.26     \\[0.5ex]
2016-05-28   &    21.83   &    0.008   &    21.82   &    21.84   &    80   &    0.24     \\[0.5ex]
2016-05-29   &    21.75   &    0.037   &    21.69   &    21.82   &    120   &    0.23    \\[0.5ex]
2016-05-30   &    21.79   &    0.015   &    21.76   &    21.81   &    161   &    0.22    \\[0.5ex]
2016-05-31   &    21.83   &    0.064   &    21.70   &    21.91   &    196   &    0.22    \\[0.5ex]
2016-06-01   &    21.82   &    0.023   &    21.74   &    21.86   &    235   &    0.21    \\[0.5ex]
2016-06-02   &    21.81   &    0.107   &    21.55   &    21.94   &    276   &    0.21    \\[0.5ex]
2016-06-03   &    21.87   &    0.099   &    21.61   &    22.00   &    281   &    0.20    \\[0.5ex]
2016-06-04   &    21.89   &    0.112   &    21.62   &    22.01   &    276   &    0.20    \\[0.5ex]
2016-06-05   &    21.90   &    0.150   &    21.50   &    22.03   &    271   &    0.20    \\[0.5ex]
2016-06-07   &    21.74   &    0.130   &    21.47   &    21.91   &    161   &    0.19    \\[0.5ex]
2016-06-08   &    21.72   &    0.072   &    21.59   &    21.84   &    107   &    0.19    \\[0.5ex]
2016-06-09   &    21.67   &    0.065   &    21.55   &    21.77   &    62   &    0.19     \\[0.5ex]
2016-06-10   &    21.53   &    0.011   &    21.51   &    21.54   &    20   &    0.19     \\[0.5ex]
2016-06-26   &    21.94   &    0.003   &    21.93   &    21.94   &    28   &    0.21     \\[0.5ex]
2016-06-28   &    21.98   &    0.048   &    21.88   &    22.04   &    105   &    0.20    \\[0.5ex]
\end{longtable}
\end{scriptsize}
\end{center}

\end{document}